\documentclass[useAMS,usenatbib]{mn2e}

\usepackage{aas_macros}
\usepackage{graphicx}
\usepackage{amssymb}
\usepackage{amsmath}
\usepackage{color}
\usepackage{changes}
\usepackage{multirow}

\newcommand{\Ms} {$\rm{M_{\odot}}$}

\usepackage{hyperref}
\begin{document}

\title[LWBG \& streaming velocities: First star formation]
{The influence of streaming velocities and Lyman-Werner radiation on the formation of the first stars} 
\author[Schauer et al.]{Anna T. P. Schauer$^{1}$\thanks{E-mail: 	
anna.schauer@utexas.edu}\thanks{Hubble Fellow}, 
Simon C. O. Glover$^{2}$, Ralf S. Klessen$^{2,3}$, Paul Clark$^{4}$\\
$^{1}$ Department of Astronomy, The University of Texas at Austin, Austin, TX 78712, USA\\
$^{2}$ Universit\"at Heidelberg, Zentrum f\"ur Astronomie, Institut f\"ur Theoretische Astrophysik, Albert-Ueberle-Str. 2, 69120 Heidelberg, Germany\\
$^{3}$ Universit\"at Heidelberg, Interdisziplin\"{a}res Zentrum f\"{u}r Wissenschaftliches Rechnen, Im Neuenheimer Feld 205, 69120 Heidelberg, Germany\\
$^{4}$ School of Physics and Astronomy, Queen's Buildings, The Parade, Cardiff University, Cardiff, CF24 3AA }

\pagerange{\pageref{firstpage}--\pageref{lastpage}} \pubyear{2018}

\maketitle

\label{firstpage}
\begin{abstract}
The first stars in the Universe, the so-called Population III stars, form in small
dark matter minihaloes with virial temperatures $T_{\rm vir} < 10^{4}$~K. Cooling in these minihaloes is dominated by molecular hydrogen (H$_{2}$), and so Population III star formation is only possible in those minihaloes that form enough H$_{2}$ to cool on a short timescale. As H$_{2}$ cooling is more effective in more massive minihaloes, there is therefore a critical halo mass scale $M_{\rm min}$ above which Population III star formation first becomes possible. Two important processes can alter this minimum mass scale: streaming of baryons relative to the dark matter and the photodissociation of H$_{2}$ by a high redshift Lyman-Werner (LW) background. In this paper, we present results from a set of high resolution cosmological simulations that examine the impact of these processes on $M_{\rm min}$ and on $M_{\rm ave}$ (the average minihalo mass for star formation), both individually and in combination. We show that streaming has a bigger impact on $M_{\rm min}$ than the LW background, but also that both effects are additive. We also provide a fitting functions quantifying the dependence of $M_{\rm ave}$ and $M_{\rm min}$ on the streaming velocity and the strength of the LW background.
\end{abstract}

\section{Introduction}\label{introduction}
The transition from metal-free to metal-enriched star formation 
marks a fundamental period in the early Universe, taking place at redshifts $z \ge  10$. The so-called 
Population~III (Pop~III) stars are the first stellar objects that 
formed out of a gas  consisting only of hydrogen, helium, and negligible amounts of lithium, making 
their formation conditions different from present-day star  formation 
\citep{yoshida12,bromm13, glov13, klessen2019}. 

These first stars are out of reach of current and upcoming telescopes. Although their emission peaks in the rest-frame ultraviolet, this is shifted into the near-infrared at the present day because of the high redshifts of these stars. This greatly hampers efforts to detect them with ground-based telescopes, owing to the high thermal background of the atmosphere, putting them out of reach not only of current 8~m class telescopes but also the next generation of 30~m class facilities. Space-based telescopes avoid the problems posed by the atmosphere but current and near-future examples (e.g.\ HST, JWST) are too small to observe the first stars directly \citep{zack11,zack15,jeon19,schauer20}, with only the supernova explosions that mark the end of the lives of massive Pop~III stars potentially being detectable \citep{rydberg2020}.

Another method for exploring the properties of the first stars comes from 21\,cm radiation \citep{furlanetto06}. 
Before reionization, the hyperfine transition of atomic hydrogen can give insight into the global state of the  gas, which depends on the properties of the first stars and galaxies. A number of different experiments around the world are currently attempting to measure this signal \citep{singh17, fialkov18, leda18, philip19}, with a recent success reported by EDGES \citep{bow18}. Measurements of the high-redshift 21~cm signal allow us to infer a number of important quantities, such as the 
high $z$ star formation rate density, the X-ray emissivity of the first star-forming systems, or the minimum halo mass required for Pop~III formation \citep{fialkov19,mirocha19,schauer19b}. The parameter  space, however, is large, and often allows for more than one solution. 

Cosmological hydrodynamical simulations  are required 
to understand the nature of the first stars and star-forming  regions better. These simulations have been performed for over twenty years. 
From these simulations, we have learned that the first stars are  generally more massive than 
present day stars, with a clump mass of 100-1000\Ms, 
\citep{bromm99,abel02}. Some more modern simulations see fragmentation and 
multiple lower-mass cores \citep{clark11,get12,stacy14,Wollenberg2020}, with a high binary fraction 
\citep{stacy10,stacy13}, while others still find the high-mass formation mode to be the most prominent one \citep{hirano14,susa14}. 

An important open question that we can address with the help of cosmological simulations is the mass scale of the first dark matter haloes to host Pop~III star formation. There is broad agreement that the first stars formed in H$_2$-cooled dark matter ``minihaloes'', with masses of a few $10^5$\,{\Ms} - $10^7$\,{\Ms} and formation redshifts $z \sim 15$--30 (see e.g.\ \citealt{glov13}, \citealt{Schauer19}). However, star formation in these systems is strongly influenced by the strength of the high-redshift Lyman-Werner (LW) background and by the size of the relative streaming velocity between the baryons and the dark matter.

LW photons between 11.2\,eV and 13.6\,eV are produced in abundance by young massive stars and can dissociate H$_{2}$ via the two-step Solomon process \citep{field66,stecher67}. The presence of a high-redshift LW background produced by the earliest Pop~III stars results in a reduction in the amount of H$_{2}$ present in minihaloes, suppressing cooling and star formation in them to an extent that depends on the mass of the minihalo and the strength of the LW background \citep[see e.g.][]{MBA2001,oshea08, hirano15}.

The streaming velocity is a large-scale difference between the motion of the gas and the dark matter in the Universe, originating from the coupling between baryons and radiation prior to the recombination epoch \citep{th10}. It also acts to suppress star formation in the lowest mass minihaloes, shifting the onset of star formation to higher mass haloes
\citep[see e.g.][]{greif11,naoz12,mcquinn12,naoz13,hirano17,Schauer19}. 
With a one sigma velocity dispersion $\sigma_\mathrm{rms} \approx 30$\,km\,s$^{-1}$, these velocities are supersonic in most of the Universe following the decoupling of baryons and radiation during the recombination epoch. As the Universe expands, the size of the streaming velocities decreases with decreasing redshift as $\sigma_{\rm rms} \propto (1+z)$. Therefore, at the onset of Pop~III star formation ($z\approx 20$), the one sigma velocity dispersion is approximately $\sigma_\mathrm{rms} \approx 0.6$\,km\,s$^{-1}$. It is negligible in the present day Universe.

Although there have been many studies of both effects in isolation, until now there has been no direct numerical study of the combination of both effects, leaving their relative importance unclear. This is the gap that this paper attempts to fill. We carry out a series of high-resolution cosmological simulations in which we vary the strength of the LW background and the size of the initial streaming velocity. We consider three different values for the radiation field strength and four different values for the streaming, leading to 12 simulations in total. Each simulation is followed down to a redshift of $z \approx 14$, by which time each has formed several thousand minhaloes, several hundred of which are able to form stars. This provides us with a large sample of haloes in which to examine the effects of the LW background and the baryon-dark matter streaming, both in isolation and together.

Our paper is structured as follows. Section \ref{methods} presents the numerical method used for our simulations, including details of our choice of chemical network, our treatment of LW radiation and the streaming velocity, and how we identify haloes. In this section, we also describe the initial conditions we adopt for our simulations. In Section \ref{qual}, we show qualitatively 
how a LW background and streaming velocities impact the large- and small-scale 
physical properties of the early Universe. 
Section \ref{quant} is focused on the key quantity of our investigation: the typical mass for a star-forming minihalo and how this depends on the strength of the LW background and size of the streaming velocity. We also provide  highly-accurate fitting functions that allows one to easily determine this mass scale as a function of both parameters. We compare our findings to previous results in the literature in Section~\ref{discussion} and conclude in Section \ref{conclusion}.
\section{Method}\label{methods}
\subsection{Numerical method}
We carry out our simulations using the {\sc arepo} moving-mesh cosmological hydrodynamics code \citep{Springel2010}. {\sc arepo} is based on a quasi-Lagrangian approach in which the fluid is discretised with an unstructured mesh that is the Voronoi tessellation of a set of mesh-generating points that move with the flow of the gas. The flux of mass, momentum, energy etc. between cells in this mesh is determined  by solving the Riemann problem at each interface between mesh cells. These can be refined according to various criteria simply by adding additional mesh generating points and appropriately partitioning the gas between the newly-created cells. The refinement criteria used in our study are discussed in Section~\ref{sec:refine} below. Dark matter is represented in the simulations by discrete collisionless particles for which the gravitational forces are computed using a 
\citet{BH86} oct tree. The gravitational force is softened as described in \citet{Springel2010}, with a fixed softening length $l_{\rm soft} = 20$~comoving~pc for the dark matter and an adaptive approach that scales with the size of the mesh cells for the gas.

The version of {\sc arepo} that we use in our study includes a model for primordial gas chemistry and cooling. Our chemical network tracks the non-equilibrium abundances of 9 species (H, H$^{+}$, H$_{2}$, D, D$^{+}$, HD, He, He$^{+}$ and e$^{-}$) as well as the equilibrium abundances of the H$^{-}$ and H$_{2}^{+}$ ions. It is based on the network used in \citet{Schauer19}, but has been supplemented with a treatment of the effects of a soft ultraviolet background. We assume that below 13.6~eV, this background has the spectral shape of a $10^{5}$~K blackbody, as would be the case if it were produced purely by massive Pop~III stars. Above 13.6~eV, we set the strength of the radiation field to zero to account for absorption by the intergalactic medium. The overall photon flux is quantified by $J_{21}$, the mean specific intensity just below the Lyman limit in units of $10^{-21} \: {\rm erg \, s^{-1} \, cm^{-2} \, sr^{-1} \, Hz^{-1}}$. The influence of the radiation background on the gas chemistry is accounted for by including the following reactions in our chemical network
\begin{eqnarray}
{\rm H^{-} + \gamma} & \rightarrow & {\rm H + e^{-}}, \\
{\rm H_{2}^{+} + \gamma} & \rightarrow & {\rm H + H^{+}}, \\
{\rm H_{2} + \gamma} & \rightarrow & {\rm H + H}, \\
{\rm HD + \gamma} & \rightarrow & {\rm H + D},
\end{eqnarray}
with $J_{21}$-dependent rates taken from \citet{GJ07} (for H$_{2}$ and HD), \citet{GS09} (for H$^{-}$) and \citet{Glover2015} (for H$_{2}^{+}$).The most important of the new reactions is the photodissociation of H$_{2}$ by LW photons in the 11.2~--~13.6~eV energy range, and so for simplicity we will often refer to our adopted radiation background as a LW background, although in reality it extends over a wider range of energies than simply the LW bands.

The rate at which H$_2$ is destroyed by LW photons is highly sensitive to the amount of self-shielding occurring in the gas \citep{Draine96}. We model the effects of H$_{2}$ self-shielding using the {\sc TreeCol} algorithm \citep{Clark12}. This uses information on the H$_{2}$ mass of each Voronoi cell and  tree node stored in the Barnes-Hut tree to compute  an approximate map of the H$_2$ column densities surrounding each cell, pixelated using the {\sc healpix} algorithm \citep{Healpix05} with $N_{\rm pix} = 48$ equal-area pixels.  In this study, we use the modified version of the {\sc TreeCol} algorithm introduced by \citet{Hartwig15a} that also accounts for the velocity of the gas.  In this version of the algorithm, the H$_{2}$ in any given node or cell only contributes to the column density map if the centre-of-mass velocity of the node or cell, ${\bf v}_{\rm com}$, satisfies
\begin{equation}
\left| {\bf v}_{\rm com} - {\bf v}_{\rm curr} \right| \leq 1.694 \, v_{\rm th},
\end{equation}
where $v_{\rm curr}$ is the velocity of the current cell (i.e.\ the one for which we are computing the column density map), and where $v_{\rm th}$ is the thermal velocity of the H$_{2}$ in that cell. 
This modification to {\sc TreeCol} accounts for the fact that H$_2$ that is highly red or blue-shifted relative to the cell of interest will not contribute to the shielding of that cell, because its LW absorption lines will be significantly Doppler-shifted relative to those in the cell of interest. In low mass minihaloes with velocity dispersions $\sim v_{\rm th}$, this effect is of limited importance, but it becomes increasingly relevant as we consider more massive minihaloes with higher characteristic velocity dispersions.

Once we have the {\sc TreeCol}-derived H$_{2}$ column density map for each cell, we can then compute the effective self-shielding factor using a self-shielding function from
\citet{Draine96}
\begin{equation}
f_{\rm sh, eff} = \frac{1}{N_{\rm pix}} \sum_{i = 1}^{N_{\rm pix}} \frac{0.965}{(1 + x_{i} / b_{5})^{2}} + \frac{0.035}{y_{i}} \exp \left[-\frac{y_{i}}{1180} \right],
\end{equation}
where $x_{i} = N_{\rm H_{2}, i} / 5 \times 10^{14} \: {\rm cm^{-2}}$, with $N_{\rm H_{2}, i}$ being the H$_{2}$ column density in pixel $i$ of the column density map, $y_{i} = (1 + x_{i})^{0.5}$, and where $b_{5} = b / 10^{5} \, {\rm cm \, s^{-1}}$ is the scaled Doppler parameter of the molecular hydrogen in the cell, which we assume to be dominated by thermal broadening. The H$_{2}$ photodissociation rate in the cell is then simply given by
\begin{equation}
R_{\rm diss} = f_{\rm sh, eff} R_{\rm diss, thin},
\end{equation}
where $R_{\rm diss, thin} = 1.38 \times 10^{-12} \, J_{21} \: {\rm s^{-1}}$ is the value in the optically thin limit.

The \citet{Draine96} self-shielding function assumes that H$_{2}$ is rotationally cold (i.e.\ that all of the molecules are in their ortho or para ground states). This is a good approximation at the low densities that we are primarily concerned with in this study, but may over-estimate the effectiveness of self-shielding in warm, dense gas \citep{WHB11,WH19}. However, we do not expect this to impact our results, since in order for the gas to collapse to the densities at which the \citet{Draine96} function becomes inappropriate, it must already have formed enough H$_2$ to provide effective cooling.

\subsection{Initial conditions}
The basic setup of our simulations is the same as in \citet{Schauer19}. Briefly, we adopt a $\Lambda$CDM cosmological model, with parameters $h = 0.6774$, $\Omega_{0} = 0.3089$, $\Omega_{\rm b} = 0.04864$, $\Omega_{\Lambda} = 0.6911$, $n = 0.96$ and $\sigma_{8} = 0.8159$ \citep{Planck2016}. The simulations are initialised at a redshift $z = 200$ in a box of size $1 h^{-1} \, {\rm Mpc}$ in comoving units. Initial conditions for the dark matter are generated using {\sc music} \citep{hahn11} with the transfer function of \citet{Eisenstein98}.  The density distribution of the baryons is assumed to initially trace that of the dark matter. This approach is not entirely correct for runs with non-zero streaming velocities \citep[see e.g.][]{naoz05,naoz09,naoz11}, but the impact of this simplification on our results should be small (see e.g.\ the discussion in \citealt{Schauer19} or the recent results of \citealt{Park20}). In runs without streaming, the velocity field of the baryons is the same as that of the dark matter. In the runs with streaming, this is included as a constant velocity offset, arbitrarily chosen to be in the $x$-direction. Dark matter is represented by $1024^{3}$ particles with a mass of $M_{\rm dm} = 99 \: {\rm M_{\odot}}$. The gas is initially distributed across $1024^{3}$ Voronoi mesh cells, each of which has a mass $M_{\rm cell} \simeq 19 \: {\rm M_{\odot}}$. In the subsequent evolution we allow for further mesh refinement in regions where gravitational contraction sets in, as discussed in more detail in Section\ \ref{sec:refine}. In \citet{Schauer19} we demonstrated that this resolution is high enough to yield converged results for the critical minihalo mass required for efficient H$_{2}$ cooling. 

\begin{table}
\caption{List of simulations performed\label{tab:sims}}
\begin{tabular}{lcc}
\hline
Name & $v_{\rm str} / \sigma_{\rm rms}$ & $J_{21}$ \\
 \hline
 v0\_lw0 & 0 & 0.0 \\
 v0\_lw-2 & 0 & 0.01 \\
 v0\_lw-1 & 0 & 0.1 \\
 v1\_lw0 & 1 & 0.0 \\
 v1\_lw-2 & 1 & 0.01 \\
 v1\_lw-1 & 1 & 0.1 \\
 v2\_lw0 & 2 & 0.0 \\
 v2\_lw-2 & 2 & 0.01 \\
 v2\_lw-1 & 2 & 0.1 \\
 v3\_lw0 & 3 & 0.0 \\
 v3\_lw-2 & 3 & 0.01 \\
 v3\_lw-1 & 3 & 0.1 \\
\hline
\end{tabular}
\end{table}

The initial chemical state of the gas in our simulations is specified in Table~\ref{tab:chem}. The fractional abundances listed there are defined as $x_{i} = n_{i} / n$, where $n_{i}$ is the number density of the chemical species and $n$ is the number density of hydrogen nuclei. 

\begin{table}
\caption{Initial chemical abundances \label{tab:chem}}
\begin{tabular}{lc}
\hline
Species & Fractional abundance \\
\hline
e$^{-}$ & $10^{-4}$ \\
H$^{+}$ & $10^{-4}$ \\
H$_{2}$ & $2 \times 10^{-6}$ \\
H & 0.9999 \\
D$^{+}$ & $2.6 \times 10^{-9}$ \\
HD & 0.0 \\
D & $2.6 \times 10^{-5}$ \\
He & 0.079 \\
He$^{+}$ & 0.0 \\
\hline
\end{tabular}
\end{table}

In the study presented in this paper, we vary two free parameters: the strength of the Lyman-Werner background, as quantified by $J_{21}$, and the size of the streaming velocity. For the strength of the Lyman-Werner background, we consider three values: no background (i.e.\ $J_{21} = 0$), $J_{21} = 0.01$ and $J_{21} = 0.1$. Our choice of these particular non-zero values is motivated by the fact that we expect them to bracket the average value present in the Universe at redshifts $1 + z \sim 15$ and above (see e.g.\ the models for the Lyman-Werner background presented in \citealt{Haiman2000}, \citealt{trenti09}, \citealt{ahn09}, \citealt{wise12} or \citealt{agarwal12}, all of which predict values for $J_{21}$ between these limits for redshifts $15 < 1 + z < 25$ for a broad range of input parameters). Although we expect $J_{21}$ to be higher at redshifts closer to the redshift of reionization, or in the immediate vicinity of massive star-forming galaxies \citep{ahn09}, these values are representative of the ones seen by most Pop~III star-forming minihaloes during the range of redshifts in which Pop~III star formation dominates. We also note that previous work \citep[e.g.][]{Haiman2000,MBA2001,yoshida03,oshea08} suggests that these field strengths are large enough to yield significant suppression of H$_2$ cooling in high-redshift minihaloes. 

For the size of the streaming velocity, we consider values $v_{\rm str} = 0, 6, 12$ and $18 \, {\rm km \, s^{-1}}$ at $z=200$, corresponding to 0, 1, 2 and 3 times $\sigma_{\rm rms}$, the root-mean-squared streaming velocity at that redshift. The combinations of parameters that we examine are summarised in Table~\ref{tab:sims}. 

In the runs with non-zero $J_{21}$, we represent the redshift dependence of the background with a simple step function,
\begin{equation}
J_{\rm LW} = \left \{ \begin{array}{lr}
0 & \hspace{1in} \mathrm{for~} z > 24 \;,\\
10^{-21} J_{21} &  \mathrm{for~} z \leq 24\;,
\end{array} \right .
\end{equation}
where $J_{\rm LW}$ is the mean specific intensity of the LW background in units of ${\rm erg \, s^{-1} \, cm^{-2} \, Hz^{-1} \, sr^{-1}}$. Although somewhat unrealistic, this choice allows for a much cleaner numerical experiment than would be the case for a more complex dependence on $z$.

\subsection{Mesh refinement and sink particles}
\label{sec:refine}
By default, {\sc arepo} attempts to ensure that the gas mass in each Voronoi mesh cell remains within a factor of two of its initial value, which in our case is $M_{\rm cell} \simeq 19 \: {\rm M_{\odot}}$. If necessary, it will refine the mesh cells to ensure that this condition is met by adding additional mesh-generating points. We supplement this default refinement scheme with an additional Jeans criterion, and refine the grid as necessary to ensure that the local Jeans length is always resolved with at least 16 cells \cite[for further details, see e.g.][]{federrath2010}.

For efficiency reasons, we do not allow gravitationally collapsing gas to reach arbitrarily high densities. Instead, we replace regions of dense, collapsing gas with non-gaseous sink particles. The sink particle implementation we use here is the same as that introduced in \citet{Tress2020} and \citet{Wollenberg2020}. Mesh cells denser than our sink particle creation threshold, $n_{\rm th} = 10^{5} \: {\rm cm^{-3}}$, are candidates for sink particle formation. However, in order to actually form a sink, the gas in the cell and its surroundings must pass an additional series of checks. It must be gravitationally bound and collapsing (as assessed based on the divergence of the local velocity and acceleration of the gas). It must also be located at a local minimum in the gravitational potential. Finally, sink formation is disabled in gas cells located within the accretion radius of an existing sink. 

Once formed, sink particles can accrete gas from their surroundings.  Gas within a given sink accretion radius $r_{\rm acc}$ and above the threshold density $n_{\rm th}$ will be accreted by the sink provided that it is gravitationally bound to the sink particle. We adopt $r_{\rm acc} \approx 1$\,physical pc and $n_{\rm th} = 10^5$\,cm$^{-3}$, because these values roughly reflect the bulk properties of the star-forming central region of the halo \cite[e.g.][]{bromm99,abel02}, and because we do not need to follow the full fragmentation process and the formation of individual stars in the current investigation. This would require resolving much smaller scales and higher densities \citep[e.g.][]{haemmerle2020}.
As in  \citet{Tress2020} and \citet{Wollenberg2020}, we do not remove all of the gas from cells satisfying this criterion, but instead simply remove enough gas to reduce the cell density to $n_{\rm th}$. The mass and momentum of the accreted gas are added to the sink mass and momentum.

\subsection{Halo selection}
Our haloes are chosen the same way as in \cite{Schauer19}. We use the standard friends-of-friends algorithm to identify dark matter haloes. Gas cells are included as secondary components. A subfind algorithm \citep{subfind} is run in addition,  which we use to determine the centre of the halo. 

At every snapshot, we check all minihaloes with masses $M_\mathrm{halo} \ge 5\times10^4$\,M$_\odot{}$ and investigate whether these haloes  fulfil our star formation criteria. These are the same as in \cite{Schauer19}. We consider a halo to be capable of forming stars if at least one gas cell has:
\begin{itemize}
    \item a number density $n \ge 10^2$\,cm$^{-2}$, 
    \item a temperature $T \le 500$\,K,
    \item a molecular hydrogen abundance of $x_{\rm H_{2}} \ge 10^{-4}$. 
\end{itemize}
In contrast to our procedure in \cite{Schauer19}, we measure the mass of the star-forming halo only in the snapshot when these criteria have been met for the first time. As all DM particles have unique IDs, we 
are able to trace haloes through the entire simulation and can build  merger-trees. 
With the help of these merger trees, we are able to identify when haloes fulfil the star formation criteria for the first time. 
If we did not use merger trees, our results would be biased to higher masses, as in the absence of feedback, almost all haloes retain their star-forming ability at lower redshifts but accumulate more mass through accretion and mergers. 

Our sink particle formation criterion is more stringent than the star formation criterion described above, as sink formation occurs at a much higher density than the value we use to flag halos capable of star formation. Consequently, in any given halo that ultimately forms stars, the star formation criterion is fulfilled before sink particle formation occurs. This prompts the question of whether the two criteria actually identify the same set of halos: do all of the halos we flag as ``capable of forming stars'' go on to form sink particles? With the merger tree, we are able to trace back star-forming halos in time and associate the sink particles with their birth halo locations and times, allowing us to address this point. A full extended analysis that focuses on the spread of the star-forming halo masses, as well as the merger history, lies beyond the scope of this paper and will be addressed in future work. Nevertheless, a comparison of the two criteria for our most representative simulation (v1\_lw-2) reveals that all all sink particles form in halos previously flagged as capable of star formation. The time delay between the fulfilment of the star formation criterion and the actual formation of a sink particle is on average one timestep, $\Delta z = 1$, and all star-forming halos flagged between redshifts $z=20$ and $z=17$ have formed a sink particles before the end of the simulation at $z=14$.

\section{From large to small scales: a qualitative analysis}\label{qual}
As discussed in the Introduction, both the streaming velocity and the LW background
suppress Pop~III star formation through different mechanisms. We start by giving 
a qualitative analysis of our simulations, and proceed by discussing the effects that these processes have on large scales (comparable to the size of the simulation box) and small scales (comparable to the size of a halo).
\subsection{The effects of a non-zero streaming velocity}
\begin{figure*}
    \centering
    \includegraphics[height=13cm]{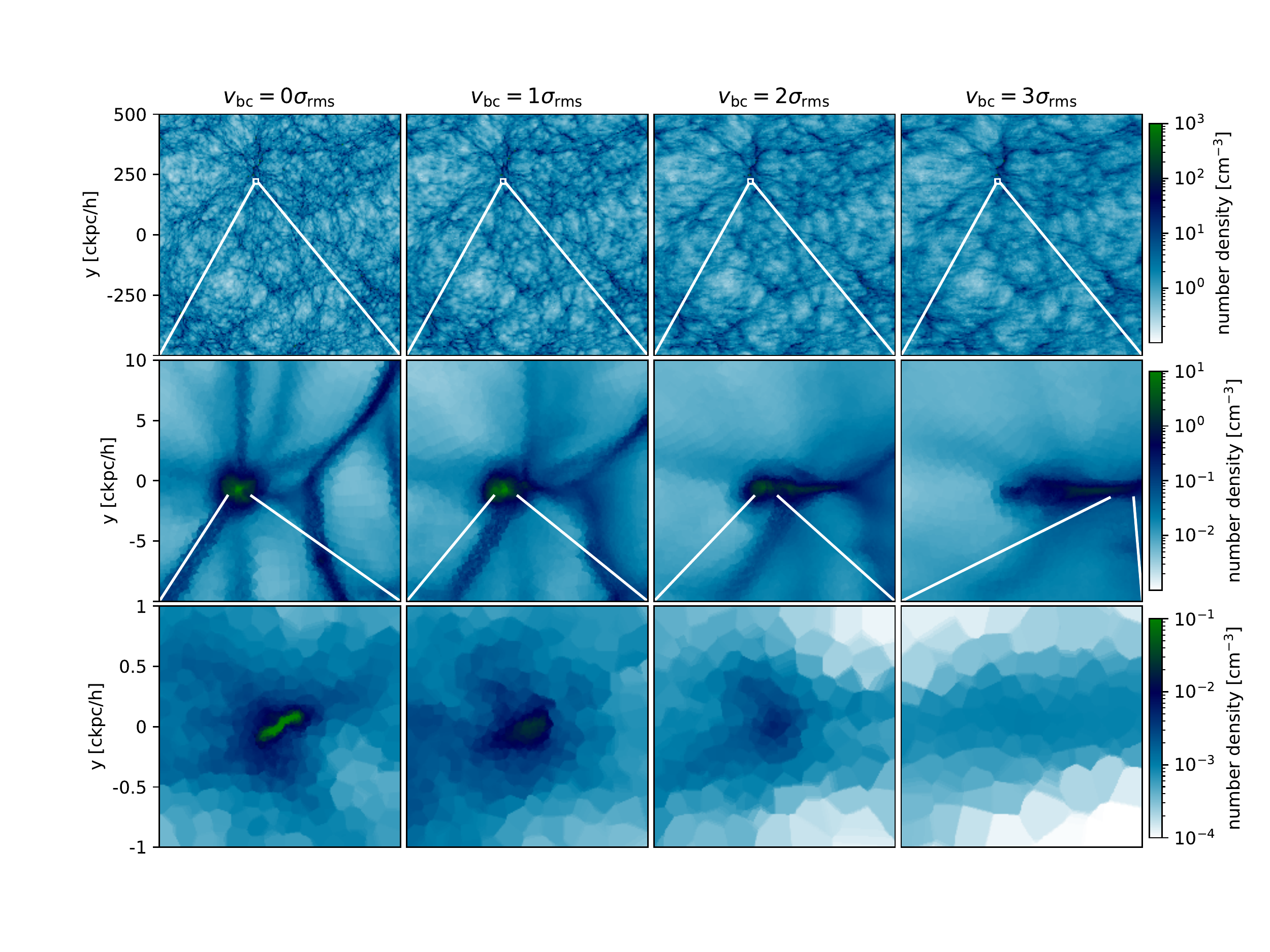}
    \caption{Gas number density slices of the four simulations without LW background at redshift $z=15$: v0\_lw0 (first column), v1\_lw0 (second column), v2\_lw0 (third column), v3\_lw0 (fourth column).  We show the whole box in the top row and a 20\,ckpc$/h$ excerpt in the middle row. In the bottom row, we show a 2\,ckpc$/h$ close-up, centred around the highest density gas from the middle row. }
    \label{fig:nslice}
\end{figure*}
In Figure \ref{fig:nslice}, we show slices of the gas number density at redshift $z=1{}5$ in four simulations with different values for the streaming velocity, progressively zooming into the centre of a halo (top to bottom row). 
Our four simulations shown here, v0\_lw0 (first column), v1\_lw0 (second column), v2\_lw0 (third column), and v3\_lw0 (fourth column), all have a zero LW background. 
In the top row, a density slice through the entire computational volume can be seen for each of these simulations. While the density structure in the image on the left is well defined and crisp, it becomes more blurred going to the right. In the rightmost panel, which shows our highest streaming velocity value of 3$\sigma_\mathrm{rms}$ the gas distribution is quite fuzzy and washed out.  The effect is especially strong in the $x$-direction, which corresponds to the direction of the streaming velocity in the initial conditions
On these large scales, one can directly observe the effect of streaming velocities: they smooth out the gas distribution. 

The middle row of Figure \ref{fig:nslice} shows a close-up with a side length of 20\,ckpc/$h$, 
centred around a minihalo (the size of the region is indicated in the top row).
For this zoom-in view, we  chose exactly the same $x$, $y{}$ and $z{}$ 
coordinates in all simulations to be able to better compare the effects of the different streaming velocities. As the streaming velocity increases, the highest density region shifts to the right, in the direction of the streaming velocity.
 
As seen on the large scales, streaming velocities lead to the washing out of the gas structures, whereas the dark matter structure remains largely unchanged.  
All structure in the Universe forms in a hierarchical manner: first, sheets and filaments emerge, before they collapse into haloes, with smaller haloes  forming first. 
For a large time interval in the early Universe, these relative velocities between gas and dark  matter were larger than the escape velocities of the dark matter potentials of the filaments and haloes that began emerging through gravitational interactions. 
As a result, the gas follows the dark matter structure later when the streaming velocity is 
higher, and we can see a lower density contrast in the slice plots 
 \cite[e.g. compare to][who find a phase shift between the linear collapse modes of gas and dark matter]{naoz14}. 
 
As a consequence of this behaviour, the gas fraction in the haloes is reduced
\citep{t10,popa16,Schauer19}, as is the mean gas density within the virial radius. 
Even the dark matter power spectrum is slightly reduced \citep{oleary12,ahn18}.

This directly impacts the ability of the gas in the haloes to cool and form stars. Haloes formed in regions of the Universe with high streaming velocities have less gas than those formed in regions with lower streaming velocities, and the gas that they do retain is less dense overall, and hence less able to form H$_{2}$ and cool. 
 This can be seen clearly in the bottom row of Figure \ref{fig:nslice}, where we show a 2\,ckpc/$h{}$ close-up on the minihalo, centred around the highest density region (indicated by the white lines in the middle row). One immediately sees that the maximum density decreases with increasing streaming velocity. As star formation requires gas to reach high densities, it is not surprising that high streaming velocities hinder and delay the process.  As we will show in the next subsection, only more massive haloes are able to retain enough high-density gas for Pop~III  star formation to proceed when the streaming velocity is large.

\begin{figure*}
    \centering
    \includegraphics[height=11cm]{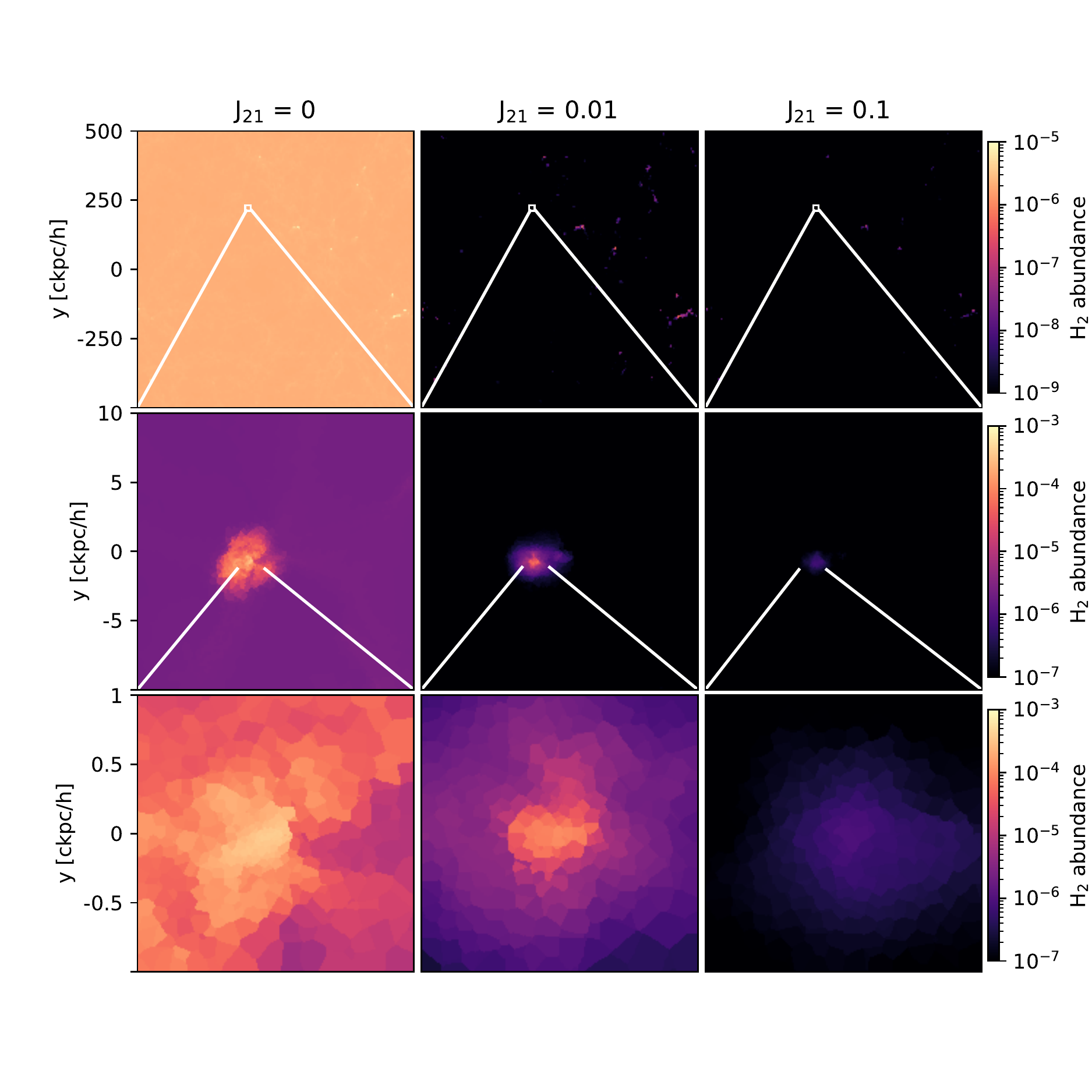}
    \caption{H$_2{}$ abundance slices of the three simulations with a streaming velocity of 1$\sigma_\mathrm{rms}{}$ and varying the LW background at redshift $z=15$: v1\_lw0 (first column), v1\_lw-2 (second column), and v1\_lw-1 (third column). We show the whole box in the top row and a 20\,ckpc$/h$ excerpt in the middle row. In the bottom row, we show a 2\,ckpc$/h$ close-up, centred around the highest density gas from the middle row. }
    \label{fig:h2slice}
\end{figure*}

\subsection{The effects of a Lyman-Werner background}

LW radiation is another way to suppress star formation. By photodissociating 
molecular hydrogen, the main coolant at high redshifts is destroyed. We therefore
investigate the abundance of molecular hydrogen in Figure \ref{fig:h2slice}. 
Similar to Figure \ref{fig:nslice}, the top row shows a slice of the entire 
simulation box, progressively zoomed into one halo in the middle and bottom rows. 
As it is the most common of the streaming velocity values investigated here (see Section \ref{subsec:average-minihalo}), we restrict our discussion to the 1\,$\sigma_\mathrm{rms}{}$ case, and show simulations with no LW background (left column), a weak LW background (middle column), and a larger LW background (right column). 

On large scales and in low density regions, molecular hydrogen is almost
immediately destroyed. In the top row of Figure \ref{fig:h2slice}, the molecular 
hydrogen abundance in the intergalactic medium drops from a few $10^{-6}{}$ in the run with no LW background to below $10^{-9}{}$ in both runs with a non-zero LW background. Molecular hydrogen, however, can self-shield, so in high-density regions, the abundance stays higher. This is illustrated by the few pink and yellow regions in the middle and right top panels. 

\begin{figure*}
    \centering
    \includegraphics[height=13cm]{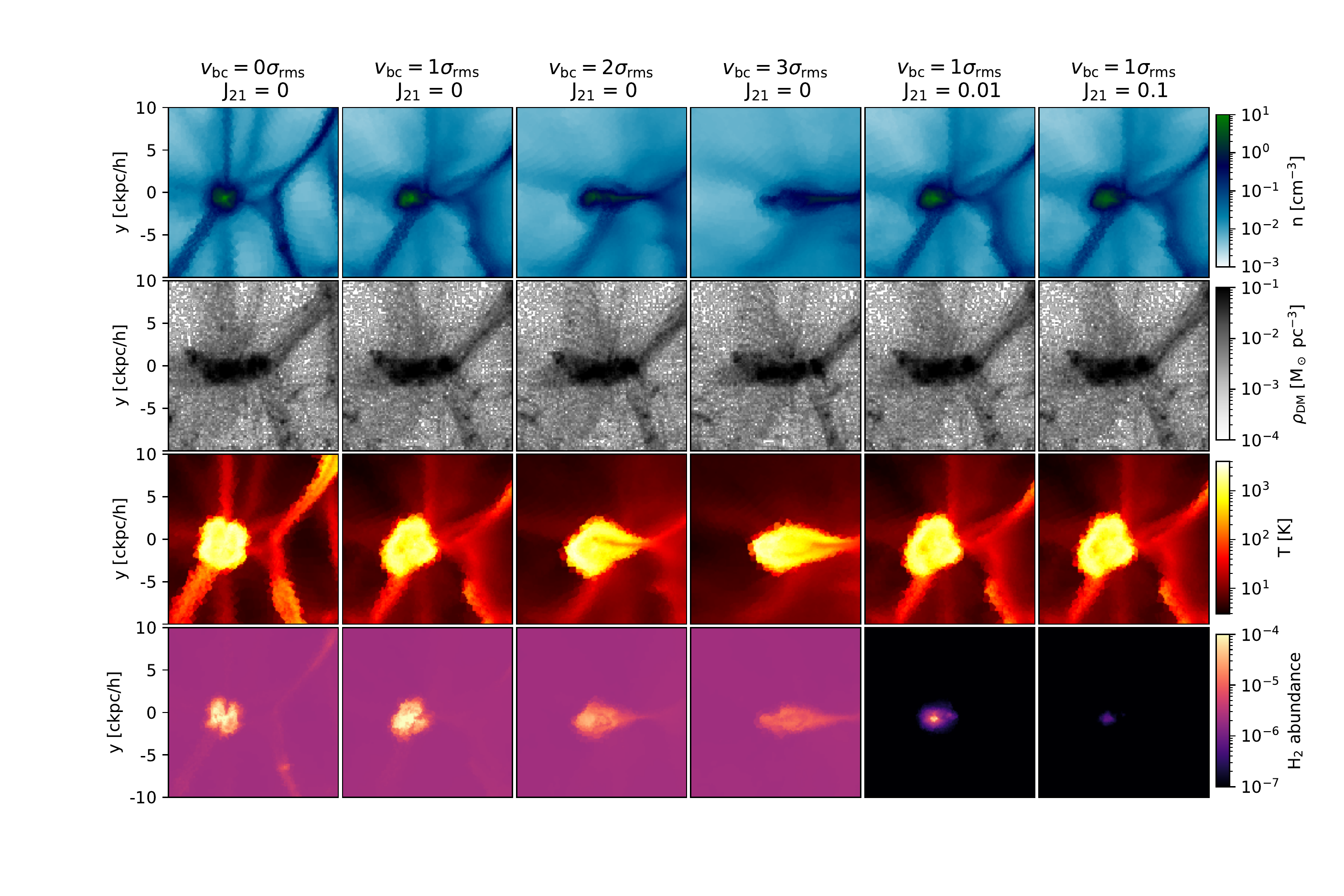}
    \caption{Slice plots of the gas number density (top row), dark matter density (second row), gas temperature (third  row) and H$_2$ abundance (bottom row) of the 20~ckpc/$h$ cut-out region shown in the middle rows of Figures \ref{fig:nslice} and \ref{fig:h2slice} at redshift $z=15$. From left to right, we show the simulations v0\_lw0, v1\_lw0, v2\_lw0, v3\_lw0, v1\_lw-2, and v1\_lw-1. While the number density and gas structure in general changes to lower maximum values for increasing streaming velocity values, it is only decreased a little bit for an increasing LW background. This is reflected in the temperature slices where the increased temperatures that indicate shock heating from the formation of the halo and filaments, largely follow the density structure. The dark matter density structure remains largely unchanged. For the two simulations with a non-zero LW background, the molecular hydrogen abundance drops to below $10^{-7}$, with only the centres of the haloes retaining a higher fraction. This is reflected in a slightly lower maximum gas number density, as seen in the top row. }
    \label{fig:allslice}
\end{figure*}

Zooming into one halo, as indicated by the white lines, one can identify 
the halo centres (compare the number density slice of these simulations 
in Figure \ref{fig:allslice}) by their increased molecular hydrogen abundances. 
In case of a zero LW background, the abundance is highest and the most extended, 
but even for the strong LW background with $J_{21} = 0.1$, an H$_2$ abundance of more than $10^{-4}{}$ is reached. This immediately demonstrates that the H$_{2}$ in this halo is able to self-shield effectively. Nevertheless, the peak H$_{2}$ abundance in the runs with a non-zero LW background is clearly smaller than in the run with no LW background, reducing its effectiveness as a coolant. The impact of this on the density distribution inside the halo can been seen in Figure~\ref{fig:allslice} (top row): the central density is slightly reduced in the runs with $J_{21} > 0$ compared to the case with no LW background (compare the second to the fifth and sixth panels).

Comparing our results here to those from the runs with high streaming but no LW, we see that the manner in which these two processes suppress cooling and star formation is quite different. Streaming reduces the gas density throughout the haloes, which has the knock on effect of making it harder for the gas to form H$_{2}$ and harder to cool once it has formed H$_{2}$. The LW background, on the other hand, has little effect on the gas density on halo scales and hence does not affect the ability of the halo to form H$_{2}$. Instead, it suppresses cooling by destroying most of the H$_{2}$ that does form, leaving less available to cool the gas.

\section{Properties of star-forming minihaloes}\label{quant}
As our next step, we move to a more quantitative analysis. Our goal is to understand 
when minihaloes form Pop~III stars and how the two effects, baryonic streaming and an elevated LW background, interact.
\subsection{Suppression of Pop~III star formation}
\begin{figure}
    \centering
    \includegraphics[width=1.\columnwidth]{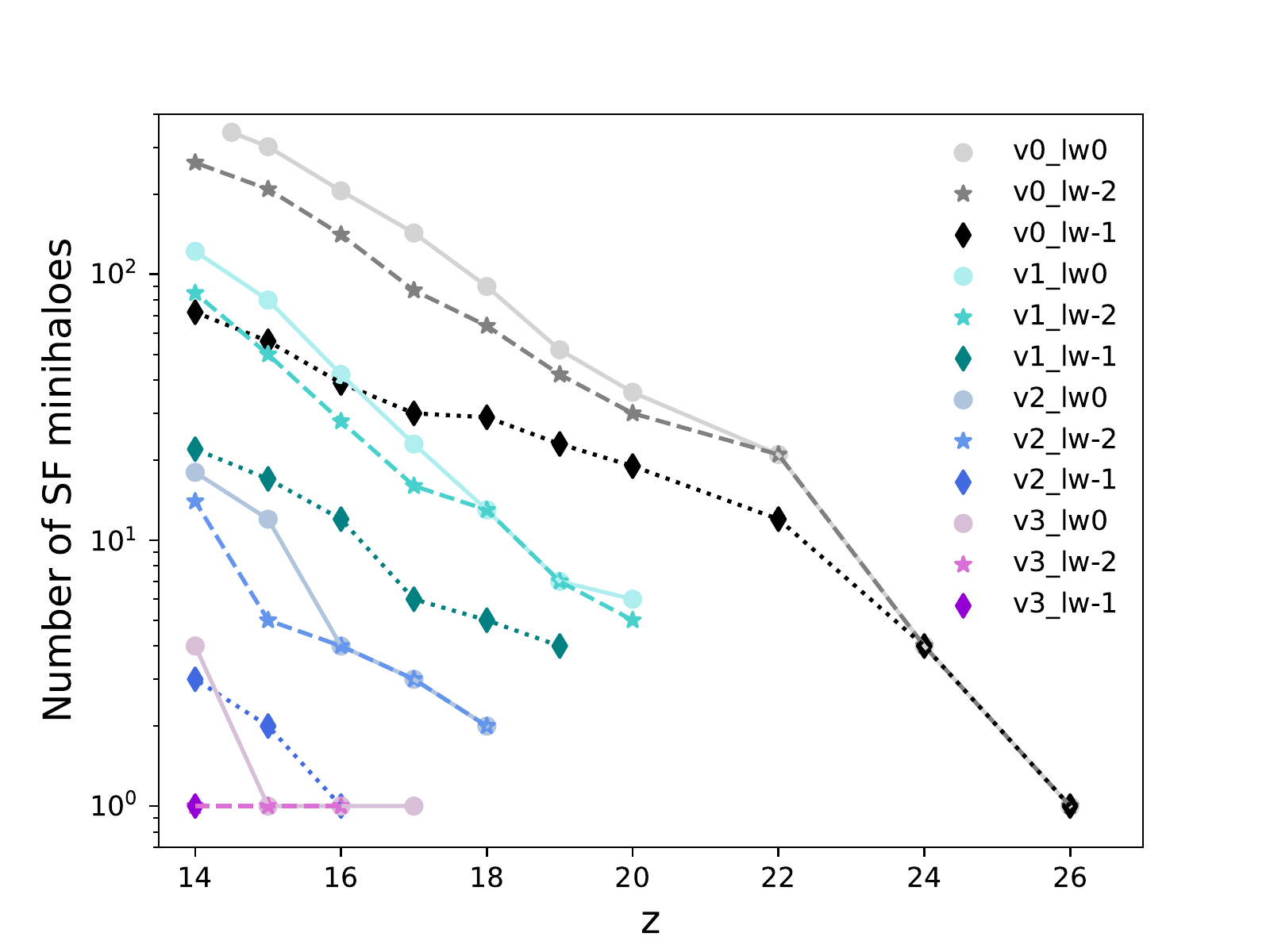}
    \caption{Number of star-forming minihaloes at each redshift for all twelve simulations. We use colours to distinguish the streaming velocity values: 
    black for 0\,$\sigma_\mathrm{rms}$, green for 1\,$\sigma_\mathrm{rms}$, 
    blue for 2\,$\sigma_\mathrm{rms}$, and pink for 3\,$\sigma_\mathrm{rms}$. 
    The LW background values are indicated by the saturation of the colour: light is used 
    for no LW background, middle for the weak background ($J_{21} = 0.01$) and dark for 
    the stronger background ($J_{21} = 0.1$). For example, light blue is simulation v2\_lw0, 
    medium blue simulation v2\_lw-2 and dark blue simulation v2\_lw-1. 
    The number of haloes increases with decreasing redshift. }
    \label{fig:nhalo}
\end{figure}

We start by investigating the number of star-forming haloes as a function of redshift
and show our results in Figure \ref{fig:nhalo}. 
As expected, the number of star-forming haloes increases with decreasing redshift, 
as haloes grow over time and the gas in the centres has more time to cool and 
collapse. 

For the first time, we show a direct comparison between the streaming velocity and LW radiation. Our key finding is that the LW background only plays a minor role in reducing the number of star-forming haloes compared to the streaming velocities. There is little difference between the results with no LW background and those for a weak background with $J_{21} = 0.01$. The presence of a weak LW background reduces the number of star-forming haloes, but only by around 25\% in most simulations. The exception is the simulation with $v_{\rm bc} = 3 \sigma_\mathrm{rms}$, where we see a larger reduction in the star-forming halo number at $z = 14$. However, the number of star-forming haloes in all of the $3 \sigma_\mathrm{rms}$ runs is so small that this could just be a consequence of small-number statistics. For the stronger LW background with $J_{21} = 0.1$, we see a larger effect: the number of star-forming haloes is reduced by a factor of 3--4 at most redshifts, leading to a significant decrease in the amount of star formation occurring in the simulation. 

Compared to this, the presence or absence of streaming velocities has a much larger impact on the number of star-forming haloes. Keeping the LW background constant, this number decreases on average by a factor of 5 for 1\,$\sigma_\mathrm{rms}{}$ compared to the zero velocity case. This reduction reaches factors of 30 -- 40 when going from zero to a streaming velocity of 2\,$\sigma_\mathrm{rms}{}$, and can exceed values of 100 when increasing the streaming velocity to 3\,$\sigma_\mathrm{rms}{}$. 

To put these numbers into context, note that the most representative value for $v_{\rm bc}$ that we examine is $v_\mathrm{bc} \approx 1\,\sigma_\mathrm{rms}$, as the volume fraction of streaming velocities peaks at 0.8\,$\sigma_\mathrm{rms}$ (compare Figure \ref{fig:volfit}), while a typical value for the LW background at the redshifts studied here is $J_{21} \approx 0.01$ \citep{wise12}.  Neglecting the effects of streaming velocities therefore has a stronger influence on the number of star-forming halos than neglecting the presence of an LW background when considering a typical region of the Universe. 
\subsection{Halo masses of star-forming minihaloes}
We investigate the halo masses of star-forming minihaloes, and by how much 
these masses are shifted to larger values for increasing 
streaming velocities and a LW background. This is a more robust measurement 
than the number of haloes that are forming stars per redshift bin, as that 
number depends on the box size and on the details of setting up the initial conditions. 

We further consider all halos only at the first snapshot (taken at equidistant redshift intervals with $\Delta z = 1$)  
in which they are forming stars. 
Pop~III stars are massive and short-lived \citep{schaerer02}, 
and one or two supernova explosions are enough to enrich the halo gas 
to metallicities of $Z \sim 10^{-3.5}$ \citep{rossi21}, 
high enough for Pop~III 
star formation to transition into Pop~II star formation \citep{bromm01}. 
By only counting halos once, we hence avoid biasing against 
higher mass halos that grew through accretion and mergers, and that 
might take Pop~II instead of Pop~III star formation into account. 

\begin{figure}
    \centering
    \includegraphics[width=1.\columnwidth]{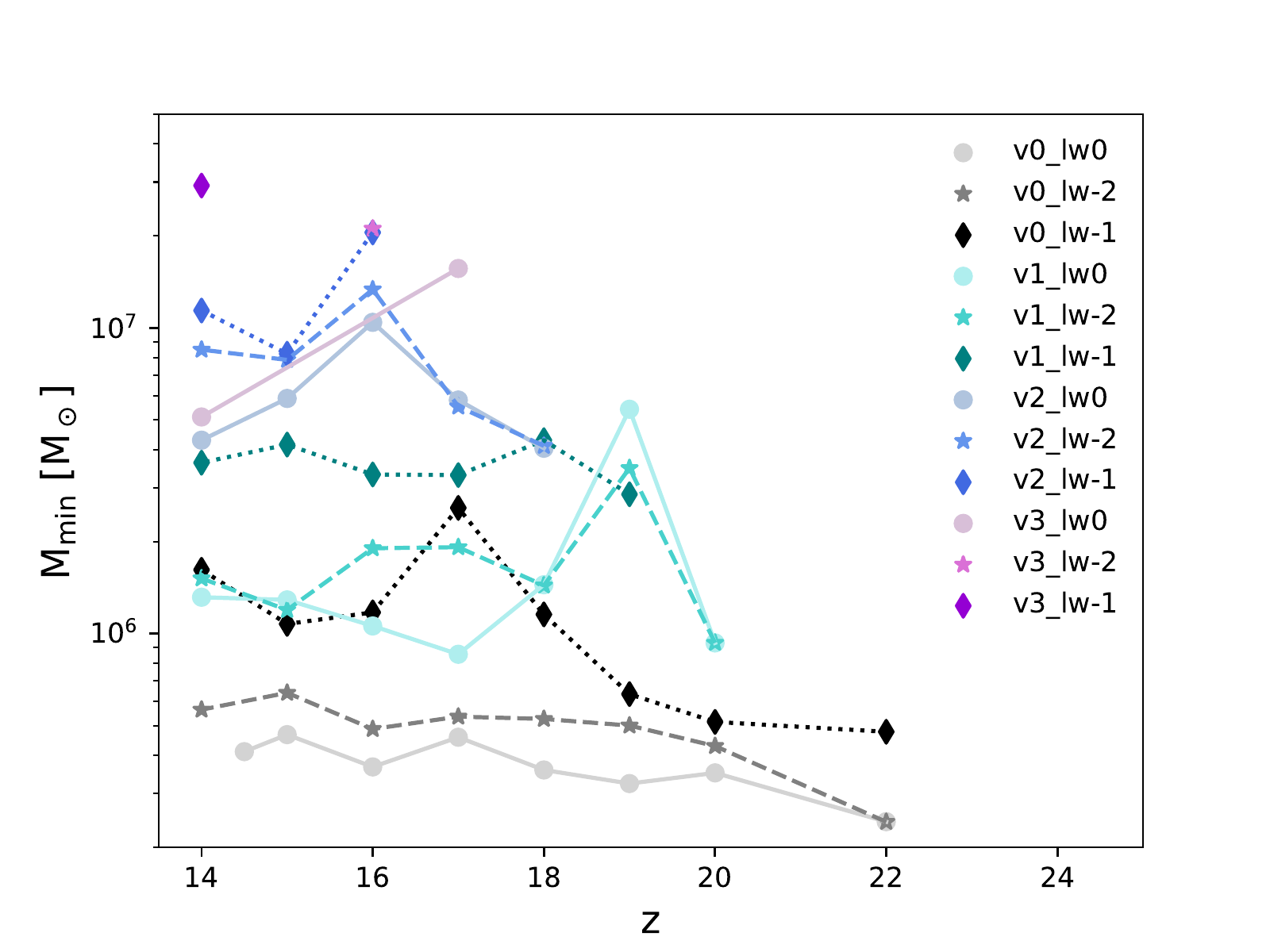}
    \caption{The minimum halo mass of star-forming haloes at the given redshift. Halos are only accounted for when they first fulfil our star formation criterion. Both streaming velocities and a LWBG lead to halos of higher masses forming 
    stars. }
    \label{fig:mmin}
\end{figure}

In Figure \ref{fig:mmin}, we show the mass of the least massive halo 
that fulfils the SF criterion as a function of redshift for all 
12 simulations. In this plot and those that follow, we  
exclude data at redshift $z\ge 24$, as we initialise 
the LW background at that redshift. If no new halo is found to be star forming at a given snapshot, we don't show a data point at this redshift. After the onset of star formation, this only happens in 
the simulations with the highest streaming velocity, 3\,$\sigma_\mathrm{rms}$.

In the simulation with no LW background and no streaming velocity, we find the lowest mass threshold: $M_{\rm min} \simeq 3\times10^5$\,M$_\odot$. Increasing the strength of the LW background increases $M_{\rm min}$, as does increasing the streaming velocity. However, once again we see that changes in the streaming velocity have a much bigger impact than changes in the strength of the LW background, at least for the range of LW background strengths considered here.

\begin{figure}
    \centering
    \includegraphics[width=1.\columnwidth]{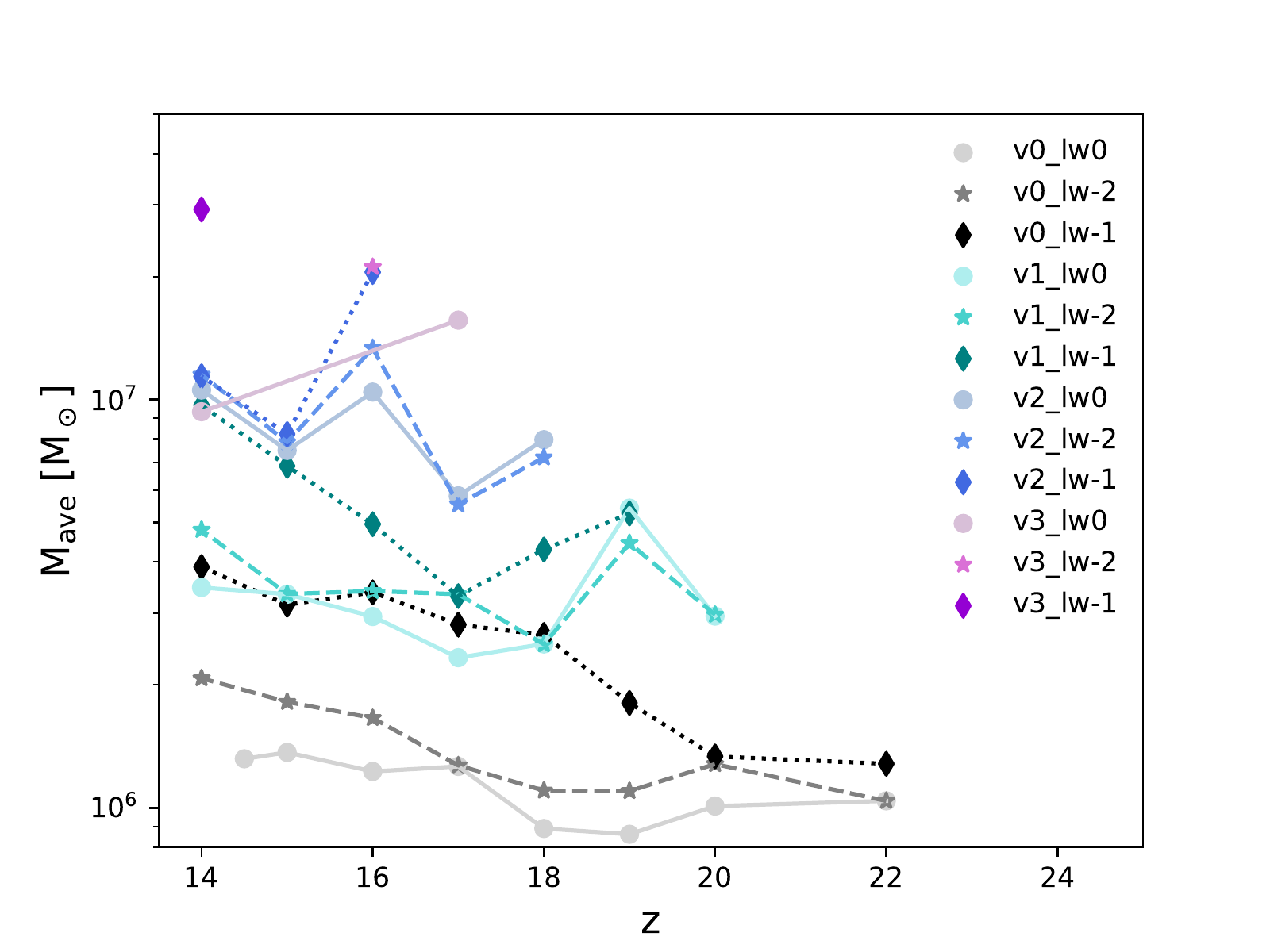}
    \caption{Same as Figure \ref{fig:mmin}, but for the average halo mass of star-forming minihaloes at that redshift. As in Figure \ref{fig:mmin}, halos are only accounted for when they first fulfil the  star formation criteria. Again, both streaming velocities and a LWBG increase the average halo mass for star formation. The behaviour is very stochastic and no clear trend with redshift can be observed. } 
    \label{fig:mave}
\end{figure}

In all of the simulations, typically only a few minihaloes with masses $M \sim M_{\rm min}$ actually form stars, while most remain starless. Star formation only becomes common in minihaloes with masses that are a factor of a few greater than $M_{\rm min}$ (see e.g.\ \citealt{Schauer19} for a more detailed discussion of this point). Therefore, as well as looking at $M_{\rm min}$, it is also useful to look at the impact of streaming and LW radiation on the average mass that a minihalo has at the point at which it first forms stars, $M_{\rm ave}$. This is plotted in Figure~\ref{fig:mave} as a function of redshift for all twelve simulations. 

\begin{figure*}
    \centering
    \includegraphics[width=1.99\columnwidth]{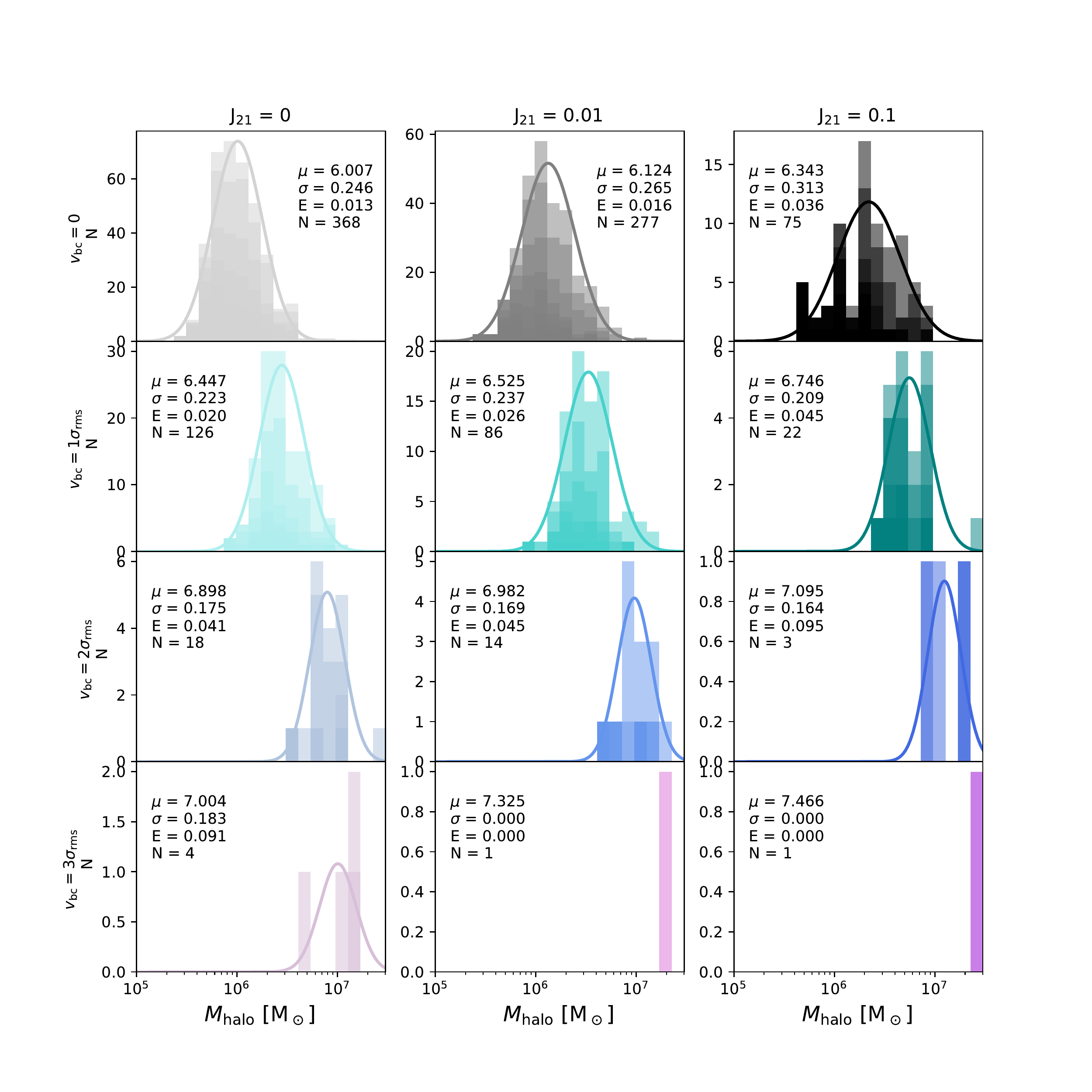}
    \caption{Histograms of the mass distributions of all star-forming minihaloes, selected at the point at which they first formed stars. We use the same colour scheme as in previous figures. In each histogram, haloes that are formed earlier 
    are shown in a darker shade of that colour. }
    \label{fig:histo}
\end{figure*}

Here again, we see that streaming velocities play a stronger role in increasing $M_{\rm ave}$ than the LW background. To better understand this behavior, we show in  Figure \ref{fig:histo} the full distribution of the masses of star-forming haloes in each simulation, selected at the point at which they first start forming stars. Note that this is not the same as selecting the haloes at a fixed point in time. In particular, the fact that many of these distributions include few or no haloes with $M > 10^{7} \: {\rm M_{\odot}}$ does not imply that haloes of this mass are unable to form stars. Rather, it implies that all haloes of this mass have at least one lower mass progenitor that was already able to form stars.

We see from the Figure that in most cases, the halo mass distributions are well fit by Gaussians. The exceptions are the runs with $3\, \sigma_{\rm rms}$ streaming, and the run with $2\, \sigma_{\rm rms}$ and $J_{21} = 0.1$, all of which produce too few star-forming haloes (typically 1--4) to let us draw any conclusions about the shape of the halo mass distribution. In each panel in the Figure, we indicate the average halo mass, $M_\mathrm{ave} = \mu$, the standard deviation of the Gaussian fit to the mass distribution, $\sigma$, the standard error in the mean, $E=\sigma/\sqrt{N}$, and the number of star-forming haloes selected from the simulation, $N$. One can see again that the average halo mass shifts to higher values for larger streaming velocities and a stronger LW background. We also note that standard deviation of the distribution is roughly 1/4 dex for the 0 and 1\,$\sigma_\mathrm{rms}$-streaming velocity simulations, but slightly smaller for higher values. This decline is unexpected and could be a result of our small sample size for large streaming velocities. A more detailed analysis requires significantly larger numerical simulations which are beyond the scope of our current investigation. 

The number of halos in our simulation is limited due to our box size 
of 1cMpc/$h$. In previous work \citep{Schauer19}, we have seen that there is 
about a one order of magnitude spread between the lest massive halo forming stars
and the most massive halo that does not form stars. Especially in simulations 
with high streaming velocity values or a LWBG of 0.1, the upper end of 
the halo mass function is influenced by our box size.

{}

\subsection{The average mass at which minihaloes first form stars}
\label{subsec:average-minihalo}
As we have already seen, neither Figure \ref{fig:mmin} nor Figure \ref{fig:mave} display a strong redshift dependence. Similar to our previous work \citep{Schauer19}, we conclude that the mass thresholds for star formation are largely independent of redshift. Therefore we can stack the data for each simulation from all redshifts in order to obtain better statistics. Our goal is to identify a general formula that links the mass of the typical star-forming halo to its environmental parameters, here the size of the streaming velocity  and the strength of the LW background it is exposed to. 

\begin{figure}
    \centering
    \includegraphics[width=1.\columnwidth]{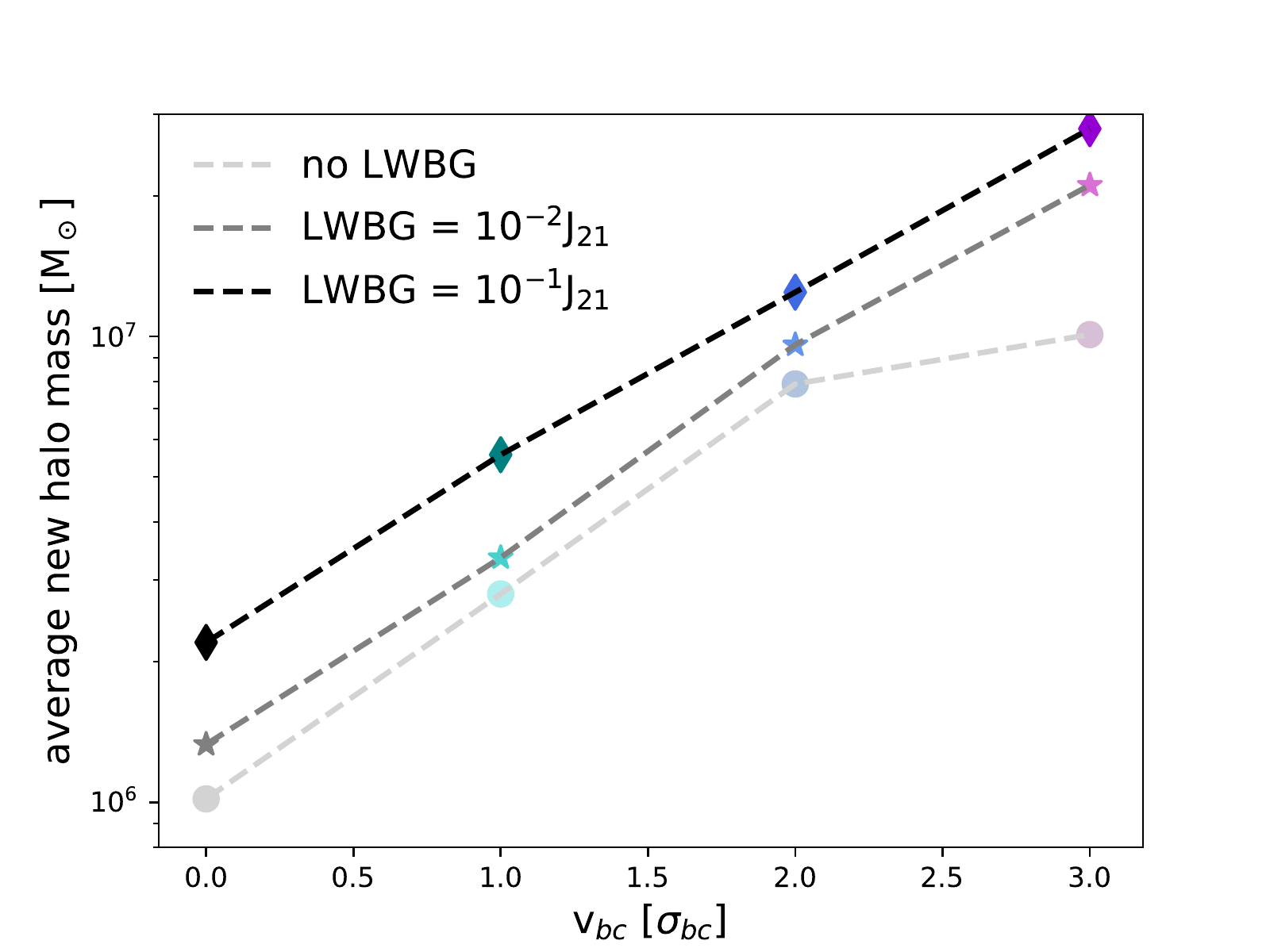}
    \caption{Average halo mass as a function of streaming velocity.}
    \label{fig:mhalo_stream}
\end{figure}

\begin{figure}
    \centering
    \includegraphics[width=1.\columnwidth]{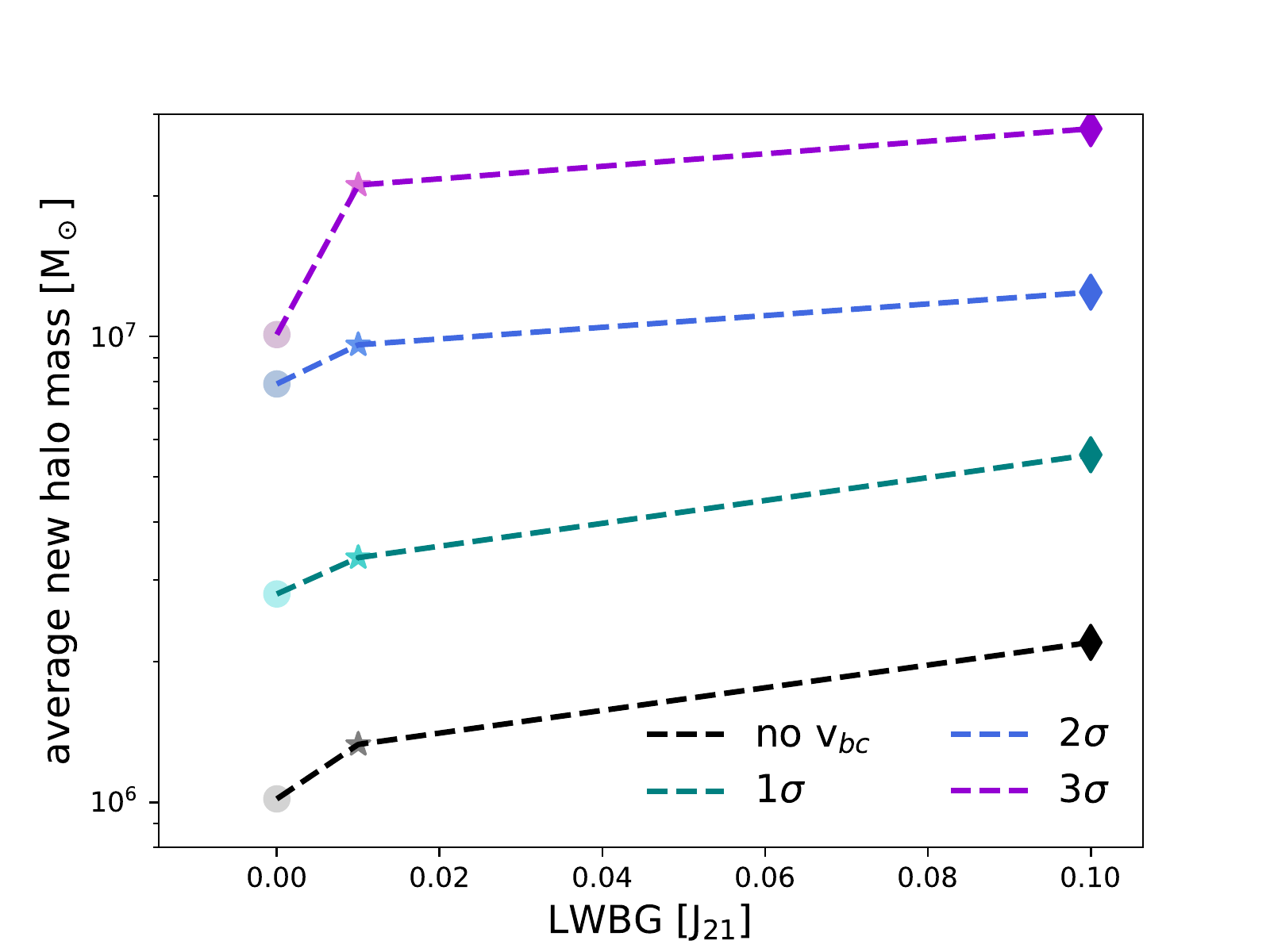}
    \caption{Average halo mass as a function of Lyman-Werner background.}
    \label{fig:mhalo_lwbg}
\end{figure}

The resulting two parameter metric is illustrated in Figure \ref{fig:mhalo_stream}, where we present the average mass at which a minihalo first forms stars as a function of the streaming velocity, and in Figure \ref{fig:mhalo_lwbg}, where we show the average halo mass as a function of the LW background. 

The average mass rises linearly with the streaming velocity, and the intercept scales as the square root of the strength of the LW background. We find that most of the data can be well fit by a relation 
\begin{equation}
\log_{10}M_\mathrm{ave} = \log_{10} M_0 + 0.4159\times \frac{v_\mathrm{bc}}{\sigma_\mathrm{rms}} ,  \label{eq:fit1}
\end{equation}
where the intercept $\log_{10} M_0{}$ is described by 
\begin{equation}
\log_{10} M_0 = 6.0174\times \left(1 + 0.166\times \sqrt{J_{21}} \right), \label{eq:fit2}
\end{equation}
and all masses are given in solar masses. 
This relation is valid for the range of values explored in our simulations, 
i.e. $0\,\sigma_\mathrm{bc} \le v_\mathrm{bc} \le 3\,\sigma_\mathrm{bc}$ 
and $0 \le J_{21} \le 0.1$. 

The minimum halo mass can be fitted with a function of the same shape. This 
time, however, the slope $s$ varies significantly with the LW background value, 
and we apply a fit to the slope as well.

\begin{equation}
    \log_{10}M_\mathrm{min} = \log_{10} M_0 + s\times \frac{v_\mathrm{bc}}{\sigma_\mathrm{rms}} ,  \label{eq:fit3}
\end{equation}
with
\begin{equation}
\log_{10} M_0 = 5.562\times \left(1 + 0.279\times \sqrt{J_{21}} \right), \label{eq:fit4}
\end{equation}
and 
\begin{equation}
s = 0.614\times \left(1 - 0.560\times \sqrt{J_{21}} \right) . \label{eq:fit5}
\end{equation}
Again, this relation is valid for the streaming velocity values and LW background 
values explored by these simulations, 
especially for $J_{21} \le 0.1$.

\begin{figure}
    \centering
    \includegraphics[width=1.\columnwidth]{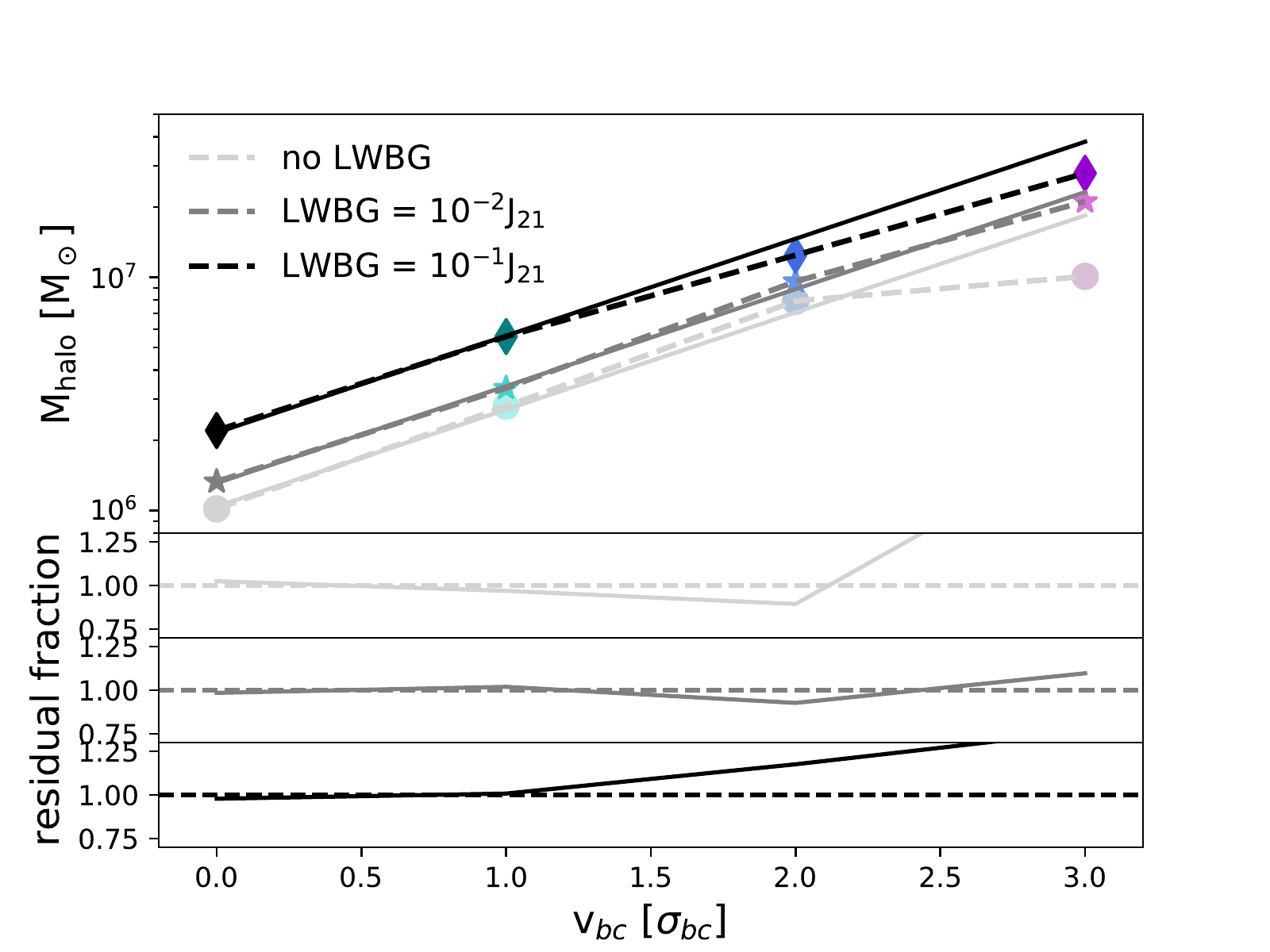}
    \caption{Average halo mass as a function of streaming velocity for our three values of the LW background. We show the data points (symbols, dashed lines) and the corresponding fit (solid lines). In the lower three panels, we show the residuals for the three values individually: no radiation (2nd panel), a LW background with $J_{21} = 0.01$ (third panel) and a LW background with $J_{21} = 0.1$ (fourth panel).}
    \label{fig:fit}
\end{figure}\textbf{}

\begin{figure}
    \centering
    \includegraphics[width=1.\columnwidth]{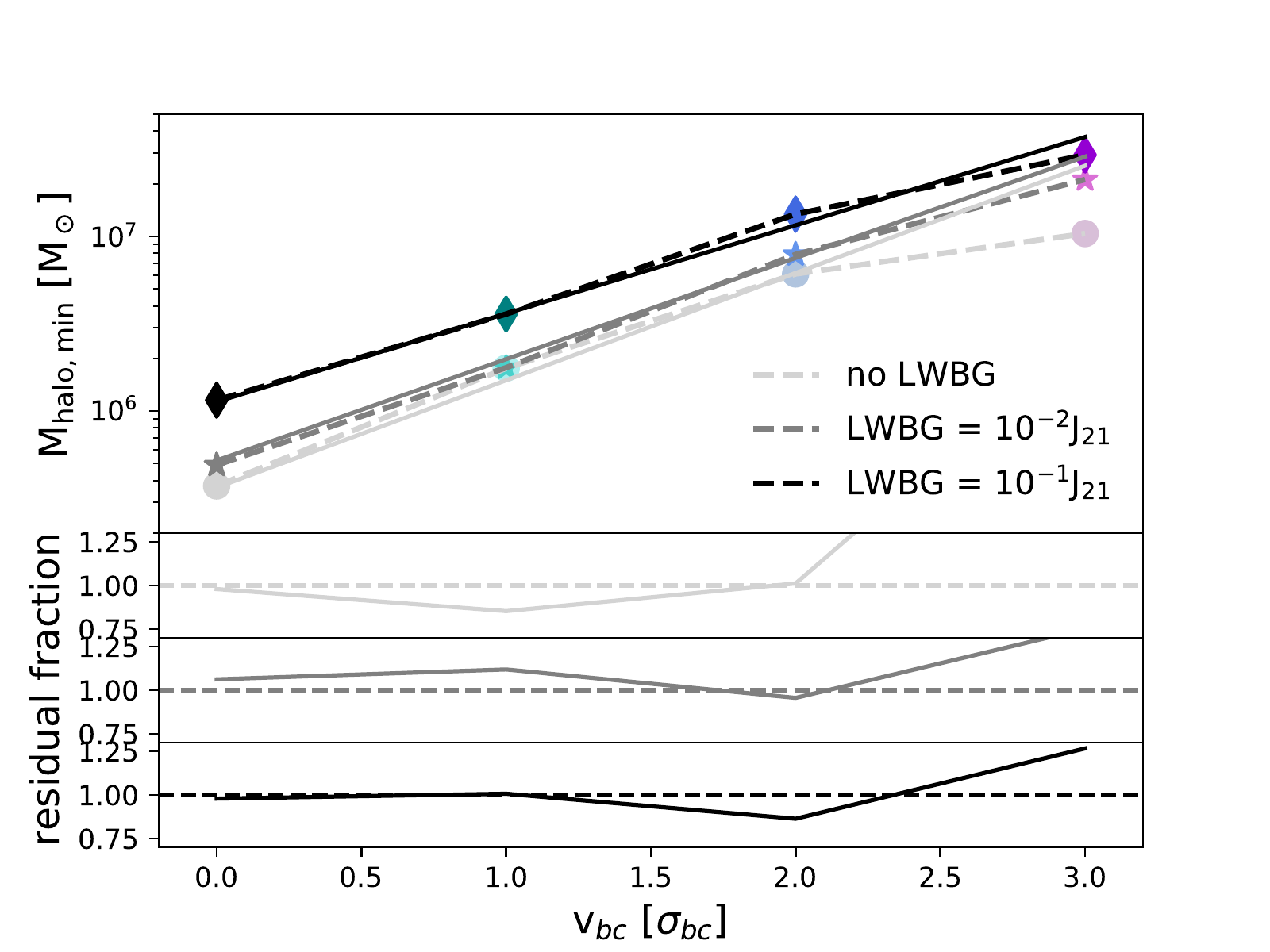}
    \caption{Same as Figure \ref{fig:fit}, but for the minimum halo mass as a function of streaming velocity for our three values of the LW background. }
    \label{fig:min-fit}
\end{figure}\textbf{}
We evaluate the fit to the data in Figures \ref{fig:fit} and \ref{fig:min-fit}, where we display the average (minimum) halo mass as a function of streaming velocity in the top panel and the residual in the bottom panel. We find that the fit (solid lines) matches the data (dashed lines) very well (within 4\%) for zero and  1\,$\sigma_\mathrm{rms}$ streaming velocities, and moderately well for 2\,$\sigma_\mathrm{rms}$ streaming. It does not fit well for the 3\,$\sigma_\mathrm{rms}$ streaming runs, but as we have already seen, these runs are strongly affected by small number statistics. 
With our fit function, the minimum halo mass exceeds the average halo mass for $v_\mathrm{bc} \approx 2.5  \sigma_\mathrm{rms}$, a result of the poor number statistics for $3 \sigma_\mathrm{rms}$.
An additional comparison of our fit functions to the data is given in the appendix, in Figures \ref{fig:fitvalues}.

\begin{figure}
    \centering
    \includegraphics[width=1.\columnwidth]{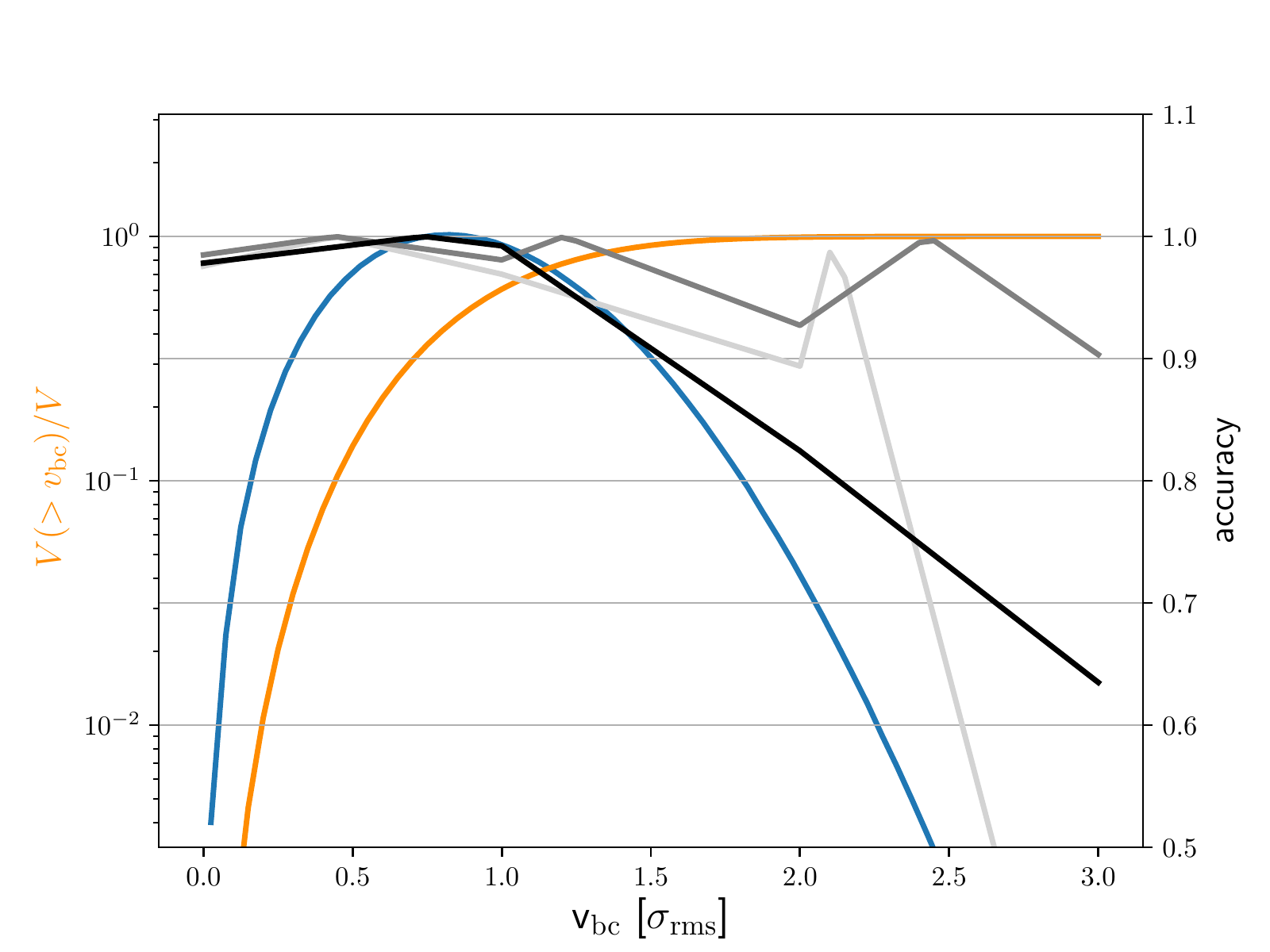}
    \caption{Differential (blue line) and cumulative (orange line) volume 
    filling fractions of the Universe, as a function of streaming velocity. The right axis shows the accuracy -- the deviation of the fit from a perfect fit -- by solid lines, for no LW background (light grey), a weak LW background with
    $J_{21} = 0.01$ (middle grey) and a strong LW background with $J_{21} = 0.1$ (black).}
    \label{fig:volfit}
\end{figure}

It is also important to note that the zero and 1\,$\sigma_\mathrm{rms}$ cases are by far the most representative of the Universe. This is quantified in Figure \ref{fig:volfit}, where we show differential (cumulative) volume filling fractions of the Universe as a function of streaming velocity, illustrated by the blue (orange) line. One can see that there is a relatively sharp peak at 0.8\,$\sigma_\mathrm{rms}$. We also include the accuracy (the deviation of the residual fraction from the perfect one-to-one fit) in this Figure as described by the light grey (a LW background with $J_{21} = 0.1$), the medium grey (a LW background with $J_{21} = 0.01$) and the black (no LW background) lines. In the low streaming velocity regions, the fit is very accurate, and overall, the fit  represents the data very well (at least 96\%). 

\textbf{}
\section{Discussion}\label{discussion}
\subsection{Comparison with previous results}
There have been a number of previous studies of the impact of a LW background on H$_{2}$ cooling in minihaloes \citep[see e.g.][]{Haiman2000,MBA2001,yoshida03,wise07,oshea08}. The majority of these studies found substantial suppression of H$_{2}$ cooling for LW background field strength $J_{21} = 0.1$, in some tension with our findings that a background with this strength has a relatively small impact. The main reason for this difference appears to be the fact that most of these previous studies neglected the effects of H$_{2}$ self-shielding, which we find to have a significant effect. Particularly informative in this regard is the study by \citet{yoshida03}. In the absence of self-shielding, they find that a field strength of only $J_{21} = 0.01$ is enough to increase $M_{\rm min}$ by more than a factor of two. On the other hand, if they account approximately for the effects of self-shielding, they find little difference between their results with $J_{21} = 0$ and $J_{21} = 0.01$, in good agreement with our findings. 

Our masses appear to be slightly larger than the values reported by \cite{hirano15}, who include a self-consistent LW background and an approximate treatment of self-shielding, but no streaming velocities, and who find haloes with virial masses of a few $10^5\,- 10^6\,\mathrm{M}_\odot$. However, it is difficult to directly compare the two simulations, as \citet{hirano15} do not report the values of $M_{\rm min}$ they obtain as a function of $J_{21}$. More recently, \citet{skinner20} report results from a simulation of minihalo formation with a strongly time-varying LW background. They find values of $M_{\rm min}$ similar to those in our study, lying mostly in the range $3 \times 10^{5} < M < 10^{6} \: {\rm M_{\odot}}$. Their model includes the effects of H$_{2}$ self-shielding and they find an even weaker dependence of $M_{\rm min}$ on the strength of the LW background than we do. However, they quantify the LW background field strength in terms of the instantaneous value at the point when a minihalo forms stars. Since their background is strongly time-varying, this can often be significantly larger than the mean value seen by the minihalo during its lifetime, making direct comparison with our time-independent values difficult.

Regarding the dependence of $M_{\rm min}$ and $M_{\rm ave}$, we find, unsurprisingly, that there is good agreement between our current results and those of our previous study \citep{Schauer19}. In that paper, we compared our results to a number of previous studies from the literature and showed that there was generally good agreement between our values of $M_{\rm min}$ and those found in previous simulations. A similar result holds for our current study. 

Finally, regarding the combined effect of the LW background and streaming, we remind the reader that our simulations represent the first comprehensive study that treats both effects simultaneously.
Between the submission of the paper and answering the referee 
report, there has been one other publication, targeting the minimum and average halo mass for star formation in minihalos \citep{kulkarni20}. 
The authors find a factor of 2-3 smaller halo masses and a mild redshift dependence. We assume the difference results from a different choice of halo mass (\citealt{kulkarni20} chose the virial mass instead of the halo mass of all 
particles and gas cells the minihalo is composed of), as well as 
an inherently redshift-dependent star formation criterion, as their 
temperature criterion for star formation in a minihalo depends on the virial temperature. The details of these difference will be explored 
in future work. 


\subsection{Other effects}
In addition to the LW background and the streaming velocity, another large-scale effect at high redshift that can potentially influence the formation of Pop~III stars is the presence of an X-ray background. X-rays will be emitted by sources such as high redshift quasars or high-mass X-ray binaries, but it is unclear whether their effect will be to enhance or delay star formation: on the one hand, they heat up their environment and therefore suppress small-scale star formation, while on the other hand, the additional ionization they provide can catalyze the formation of H$_2$ and therefore could enhance Pop~III star formation \citep{GB03,machacek03,jeon12,jeon14}. 

Ionizing radiation  can also delay star formation, if it is strong enough 
to break out from a high-redshift galaxy and affect neighbouring minihaloes. This can be the case for close pairs of low-mass haloes even at high redshift \citep{chen17}, but will become much more widespread as we approach the epoch of reionization \citep{ocvirk16}. However, both of these effects are  beyond the scope of the current paper. 

\section{Conclusions}\label{conclusion}
In this paper, we have investigated how streaming velocities and a LW background  influence Pop~III star formation in minihaloes. We find that streaming velocities play a more important role in delaying star formation and in increasing the minimum and average mass at which minihaloes first form stars that the presence of a LW background of the strength that we expect to find in the Universe during the epoch dominated by Pop~III star formation. The much higher LW radiation field strength found in the vicinity of massive star-forming galaxies (see e.g.\ \citealt{ahn09}) will likely have a much more significant effect relative to streaming, but this is relevant only for a small number of Pop~III star-forming systems and is not the common case. 

We find that, independent of redshift, the average mass at which minihaloes first form stars increases from $\sim10^6\,\mathrm{M}_\odot{}$ for no streaming to $\sim3\times10^7\,\mathrm{M}_\odot$ for $3\sigma_{\rm rms}$ streaming, i.e.\ beyond the atomic cooling threshold. We also provide two simple fitting function that describes how the minimum and the average mass vary as a function of LW background and streaming velocity, for use in future semi-analytical models or cosmological simulations.

\section*{Acknowledgments}
The authors would like to thank Volker Bromm, Mattis Magg and Omid Sameie 
for fruitful discussions, and the referee Kyungjin Ahn for careful examination of the paper and useful comments. 
Support for this work was provided by NASA through the NASA Hubble Fellowship grant HST-HF2-51418.001-A awarded  by  the  Space  Telescope  Science  Institute,  which  is  operated  by  the Association  of  Universities for  Research  in  Astronomy,  Inc.,  for  NASA,  under  contract NAS5-26555. 
The authors gratefully acknowledge the Gauss Centre for Supercomputing e.V. (www.gauss-centre.eu)
for providing computing time on the GCS Supercomputer SuperMUC at the Leibniz Supercomputing Centre
under project pr53ka. 
SCOG and RSK acknowledge financial support from the German Research Foundation (DFG) via the Collaborative Research Centre (SFB 881, Project-ID 138713538) 'The Milky Way System' (subprojects A1, B1, B2, and B8). They also thank for funding from the Heidelberg Cluster of Excellence STRUCTURES in the framework of Germany's Excellence Strategy (grant EXC-2181/1, Project-ID 390900948) and for funding from the European Research Council via the ERC Synergy Grant ECOGAL (grant 855130). They also acknowledge computing time from the Leibniz Supercomputing Centre in project pr74nu, as well as resources provided by the state of Baden-W\"urttemberg through bwHPC and the DFG through grant INST 35/1134-1 FUGG.
PCC gratefully acknowledges the support of a STFC Consolidated Grant (ST/K00926/1), as well as support from the StarFormMapper project, funded by the European Union’s Horizon 2020 research and innovation programme under grant agreement No 687528. 

\section*{Data Availability}
The data underlying this article will be shared on reasonable request to the corresponding author. 

\setlength{\bibhang}{2.0em}
\setlength\labelwidth{0.0em}
\bibliographystyle{mn2e}
\bibliography{streaming}

\appendix
\section{Goodness of the fit}
\begin{figure*}
    \centering
    \includegraphics[width=1.\columnwidth]{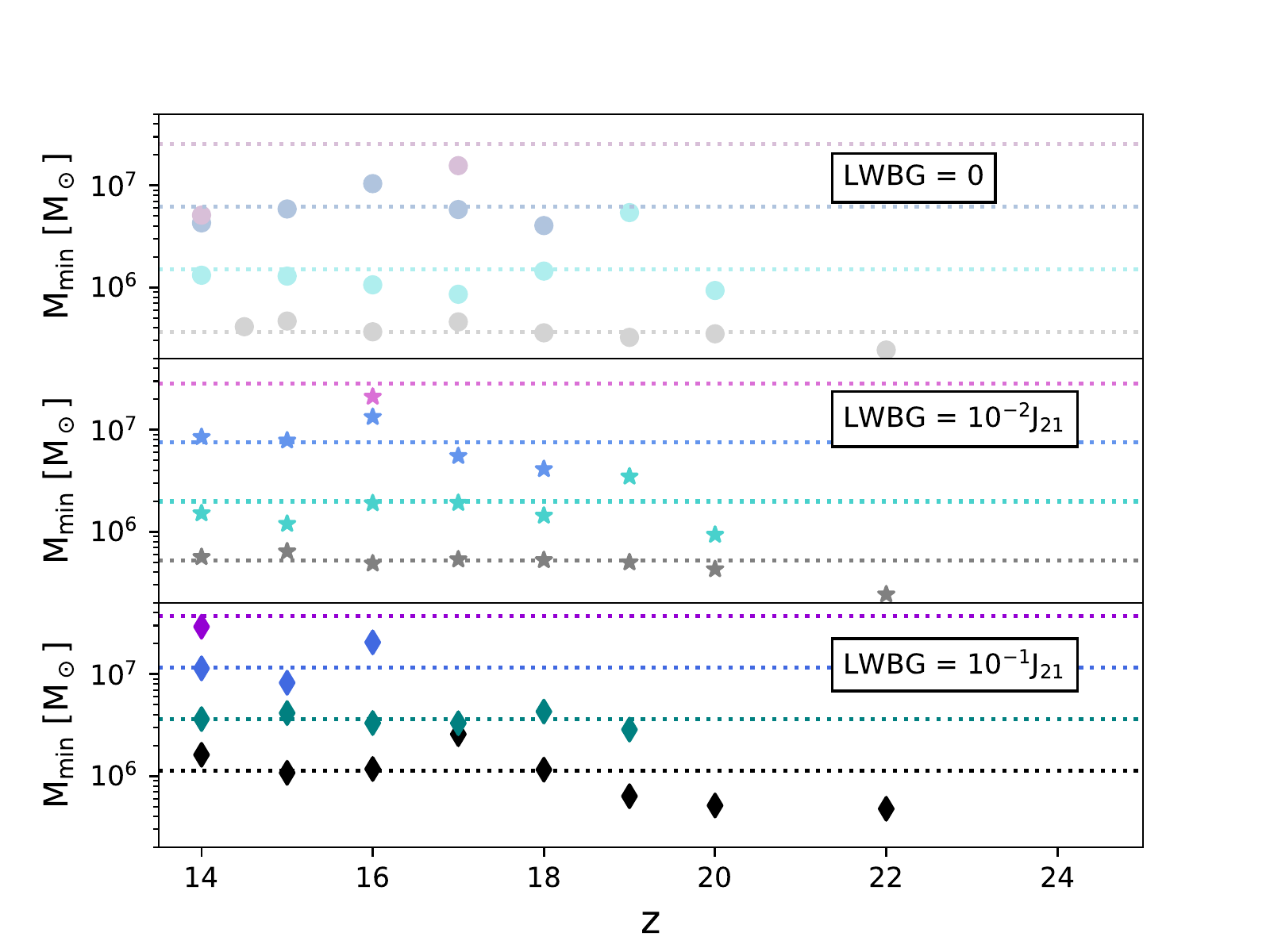}
    \includegraphics[width=1.\columnwidth]{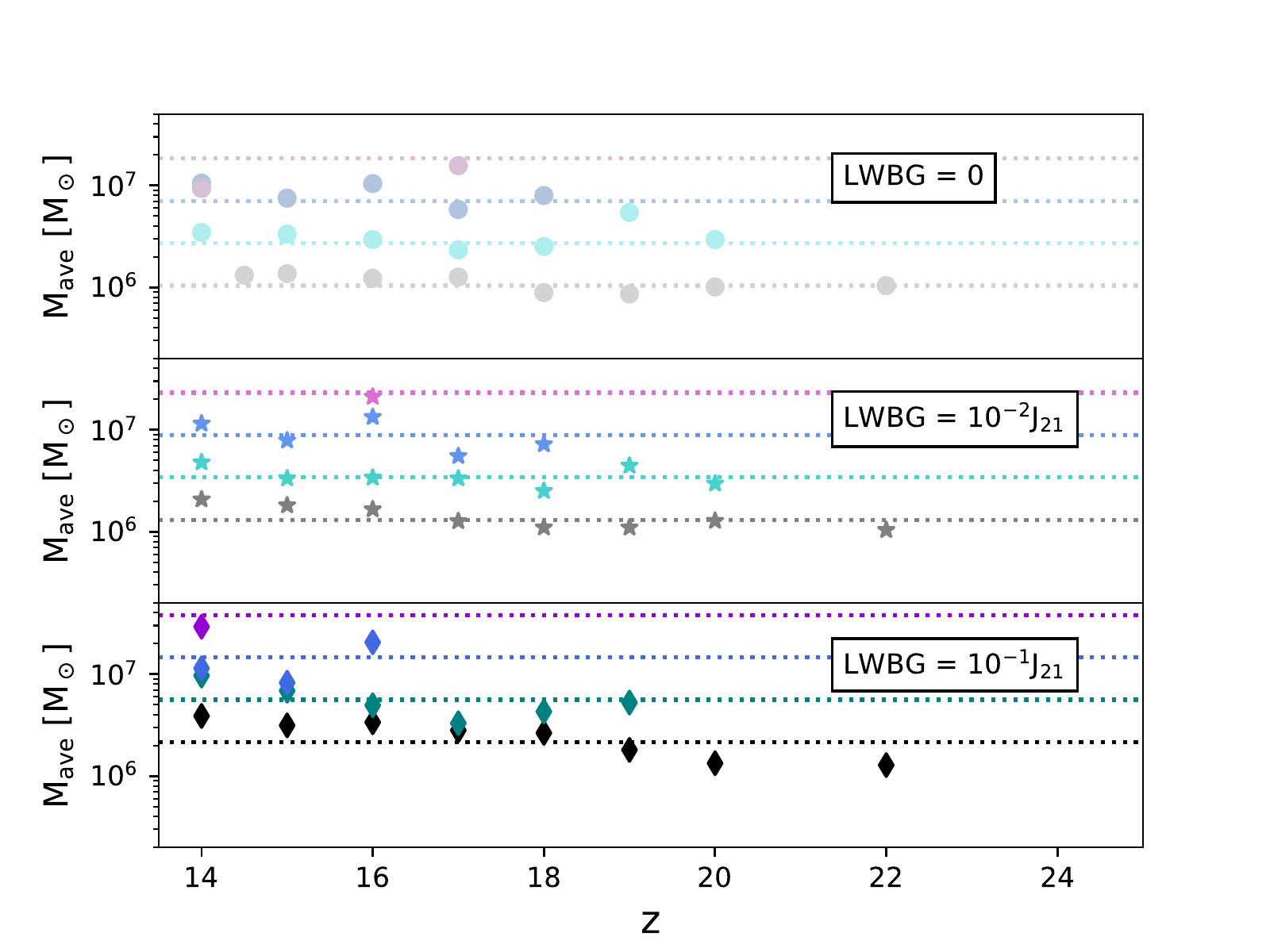}
    \caption{Minimum (left panels) and average (right panels) halo masses for star formation. We show both the data points from our simulations, in the same way we presented them in Figures \ref{fig:mmin} and \ref{fig:mave}, as well as the values obtained by our fits in Equations (9)-(13).}
    \label{fig:fitvalues}
\end{figure*}

In Figure \ref{fig:fitvalues}, we demonstrate how well our fitting functions in Equations~\ref{eq:fit1} \& \ref{eq:fit2} for the average  halo mass as well as Equations~\ref{eq:fit3}--\ref{eq:fit5} for the minimum halo mass are able to reproduce the simulation data. 

\textbf{}

\label{lastpage}

\end{document}